**Title:**

***In situ* 2D visualization of hydrogen entry into Zn-coated steels in NaCl solutions: Roles of Zn dissolution and potential distribution**

**Authors and Affiliations:**

Hiroshi Kakinuma[a,*], Saya Ajito[a,*], Koki Okumura[a,b], Makoto Akahoshi[c], Yu Takabatake[c], Tomohiko Omura[c], Motomichi Koyama[a], and Eiji Akiyama[a]

[a] *Institute for Materials Research, Tohoku University, 2-1-1 Katahira, Aoba-ku, Sendai 980-8577, Japan*

[b] *Graduate School of Engineering, Tohoku University, 6-6 Aramaki Aza-Aoba, Aobaku, Sendai, 980-8579, Japan*

[c] *Steel Research Laboratory, Nippon Steel Corporation, 20-1 Shintomi, Futtsu 293-8511, Japan*

[*] Corresponding authors. Tel: +81 22 215 2062

E-mail: hiroshi.kakinuma.a1@tohoku.ac.jp (Hiroshi Kakinuma)

E-mail: saya.ajito.d1@tohoku.ac.jp (Saya Ajito)

## Highlights:

- Hydrogen entry site was identified as the steel substrate exposed to NaCl solutions.
- Non-uniform distribution of hydrogen flux existed across the steel substrate.
- Hydrogen flux was considerably higher near the dissolving Zn coating.
- $Cl^{-}$ concentration influenced the potential distribution but had minimal effect on galvanic current.
- Hydrogen flux distribution was affected by potential distribution on the steel substrate.

## Graphical abstract

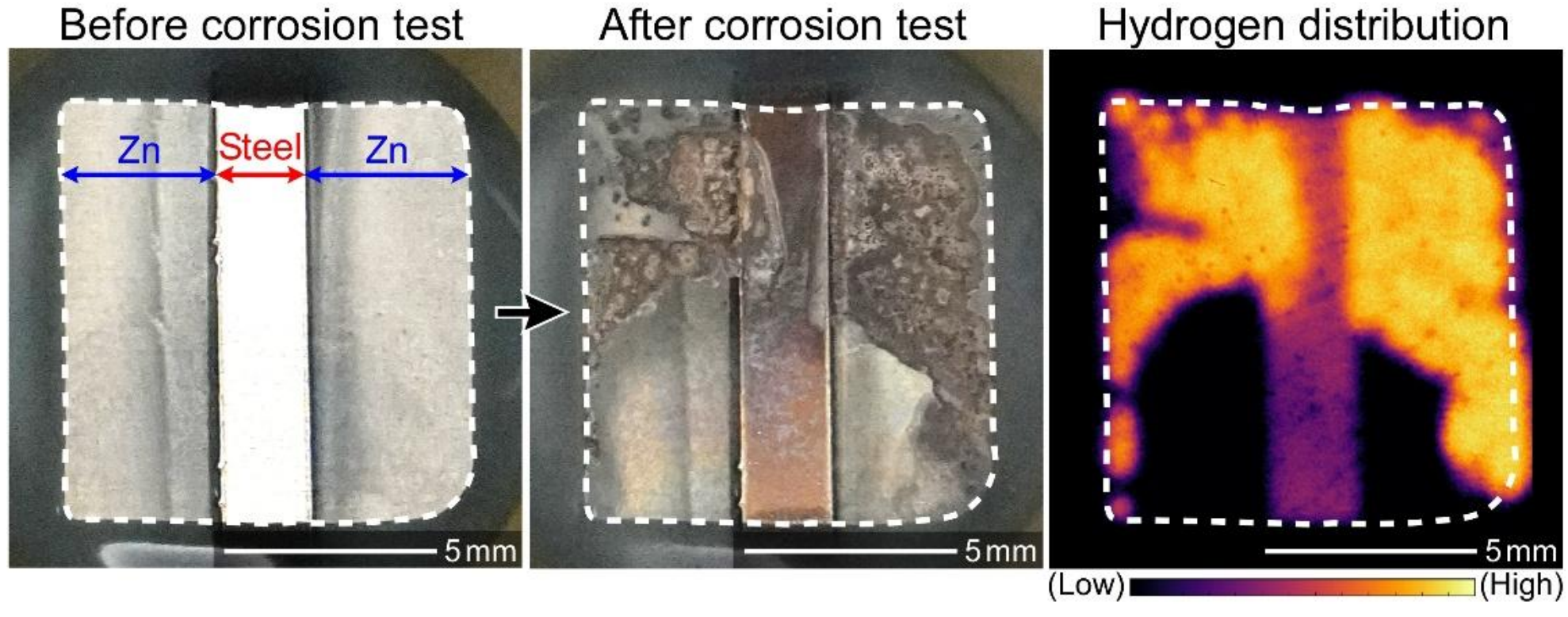

**Abstract:**

The hydrogen entry behavior of a partially Zn-coated steel sheet in NaCl solutions was investigated employing a polyaniline-based hydrogenochromic sensor, electrochemical hydrogen permeation tests, and potential measurements using a scanning Kelvin probe. While the Zn coating mitigated corrosion of the steel substrate, it simultaneously accelerated the hydrogen entry. The hydrogen entry occurred at the bare steel surface regions exposed to the NaCl solution, with the hydrogen flux exhibiting non-uniform distribution: higher near the dissolving Zn coating. While no significant differences in Zn dissolution behavior or galvanic current were observed between 0.1 and 0.01 M NaCl solutions, the total hydrogen flux decreased with decreasing $Cl^-$ concentration. This reduction was attributed to a potential gradient induced by differences in electrolyte conductivity. The results demonstrate that potential distribution, rather than galvanic current, is a dominant factor influencing hydrogen entry under the investigated conditions.

**Main text:**

## 1. Introduction

In recent years, to realize a sustainable society, global efforts have focused on mitigating climate change by reducing $CO_2$ emissions, a major contributor to environmental degradation. In the automotive industry, this prompted initiatives aimed at decreasing vehicle weight and improving fuel efficiency without compromising on crash safety [1–3]. Consequently, there is a growing demand for lightweight yet robust materials for automotive applications [1, 2]. High-strength steels, owing to their excellent strength and toughness, have therefore been widely employed as structural materials in automobiles [4, 5]. To ensure their durability in harsh atmospheric conditions, Zn coatings are commonly applied to improve corrosion resistance [6–11]. Recent studies have focused extensively on Zn-based coatings, particularly hot-dip galvanizing and galvannealing treatments [12–16]. Given their excellent mechanical properties and corrosion resistance, the application of Zn-coated steels in automotive and related industries is expected to increase in the future.

One of the primary corrosion protection mechanisms of Zn coatings is their sacrificial anodic behavior [6, 7]. The electrode potential of Zn coating is lower than that of steel, and when both are electrochemically coupled, the Zn coating serves as the anodic reaction site. This promotes cathodic reactions, such as oxygen reduction (ORR) and hydrogen evolution reaction (HER), on the steel substrate. As a result, the anodic dissolution of the steel substrate is suppressed, prolonging its service life. However, the application of Zn coatings on certain high-strength steels may enhance the risk of hydrogen embrittlement. Hydrogen embrittlement refers to the loss of ductility caused by accumulation of atomic hydrogen at trapping sites within the metal matrix [13, 17–20]. In atmospheric environments, corrosion reactions can facilitate hydrogen entry into the steel [21–25]. Hydrogen embrittlement occurs when the concentration of absorbed hydrogen exceeds a critical hydrogen concentration ($C_H$) [26]. In Zn-coated steels, exposure of the steel substrate—such as at cut edges or surface scratches—to a corrosive environment can significantly accelerate HER. This occurs because of the dissolution of the surrounding Zn coating, which facilitates hydrogen entry into the steel substrate [27, 28]. Thus, a comprehensive understanding of the hydrogen entry mechanism is essential to enable the broader application of Zn-coated high-strength steels.

The primary site of hydrogen entry on the Zn-coated steels is generally considered to be the exposed steel substrate in contact with the corrosive environment [13, 29–31]. However, the detailed characteristics of this hydrogen entry remains unclear—specifically, whether it occurs uniformly across the exposed steel substrate or exhibits a spatially varying flux distribution. Since hydrogen entry is accelerated by the dissolution of the Zn coating, it is hypothesized that the hydrogen entry is particularly enhanced in regions near the actively dissolving coating. To elucidate the fundamental mechanism of hydrogen entry, it is essential to simultaneously investigate both the dissolution behavior of the Zn coating and the corresponding hydrogen flux distribution.

$Cl^-$ concentration is a major factor influencing the corrosion of metals [32, 33]. In atmospheric environments, $Cl^-$ concentration on metal surfaces varies over time, making it crucial to evaluate its effect on service life of steels. While the influence of $Cl^-$ concentration on the corrosion behavior of Zn-coated steels has been studied extensively [34–37], its role in facilitating hydrogen entry remains largely unexplored. Elevated $Cl^-$ concentrations may enhance the galvanic current between the Zn coating and the steel substrate, resulting in accelerated hydrogen entry. Furthermore, variations in NaCl concentration alter the electric conductivity of the solution, influencing the potential distribution across the metal surface [34, 36]. Since electrochemical potential is the driving force for corrosion and hydrogen-related reactions, changes in potential distribution may affect hydrogen entry behavior. Therefore, to elucidate the hydrogen entry mechanism, it is necessary to clarify the effects of $Cl^-$ concentration on both galvanic current and potential distribution, as well as their respective roles in promoting hydrogen entry.

This study aims to elucidate the flux distribution of hydrogen entering steels (hereafter referred to as hydrogen flux distribution) and identify the dominant factors influencing hydrogen entry. To achieve this, it is necessary to visualize the two-dimensional (2D) distribution of hydrogen entry within the steel. In our previous studies [38–42], a hydrogenochromic sensor was developed using polyaniline (PANI), which undergoes a color change upon reacting with hydrogen atoms diffusing through metals [38]. Using this PANI-based sensor, the corrosion behavior of Zn-coated steels and hydrogen entry behavior were simultaneously observed *in situ*, thereby clarifying the correlation between the dissolution of Zn coating and hydrogen entry. An electrogalvanized steel was used to simplify the material factors and focus on the effect of the galvanic coupling between the Zn coating and steel substrate. Additionally, the roles of galvanic current and potential distribution in hydrogen entry were analyzed through electrochemical hydrogen permeation test, finite element method (FEM) simulations, and scanning Kelvin probe (SKP) measurements.

## 2. Materials and methods

### 2.1 Specimens

A carbon steel sheet (thickness: 0.225 mm) was cut into 30 mm × 30 mm squares and used as the specimens. The chemical composition of the steel is provided in Table I. Zn-coating was applied to one side of each specimen in a mixed solution containing $ZnSO_4 \cdot 7H_2O$ (400 g $L^{-1}$) and HCl (10 g $L^{-1}$) at 323 K. The specimens were polarized at a constant current density of $2 \times 10^3$ A $m^{-2}$ for 75 s. The estimated amount and thickness of the Zn coating were approximately 20 g $m^{-2}$ and 2.8 μm, respectively. Before the electro-galvanization, a central area of 2 mm × 30 mm was masked with tape to form a slit where the steel substrate is exposed (see Supplementary Figure 1 for schematic diagram of specimen preparation procedure). In this study, specimens with and without the Zn coating are referred to as Zn-coated and bare steels, respectively.

The bare steel surface of the specimens was mirror-polished using SiC paper and a diamond paste (6 μm, 1 μm). After polishing, the final thickness of the steel substrate was *ca.* 210–220 μm. The mirror-polished surface was then immersed in a Watts bath at 333 K and polarized at a constant current density of −30 A $m^{-2}$ for 180 s to form a Ni layer. A PANI-based hydrogenochromic sensor [38, 39, 41–43] was used to visualize hydrogen permeation through the specimens. The specimen surface covered with the Ni layer was immersed in a 0.5 M $H_2SO_4$-0.5 M aniline aqueous solution and polarized at a constant voltage of 1 V for 200 s. The intermediate Ni layer between the steel substrate and the PANI layer acts as a catalyst for the hydrogenation reaction of PANI [38]. Prior to the polymerization, electrical connections for reference and counter electrodes were made by attaching the cell cables to a Pt wire. A constant voltage of 1 V was applied between the Pt wire immersed in the solution and the specimen positioned in air. Subsequently, the specimen was immersed in the solution to initiate polymerization. This procedure served to prevent corrosion under open-circuit conditions. After polymerization, the PANI layer was rinsed with water and dried using $N_2$ gas. Prior to the hydrogen visualization tests, the specimen was stored in air for over 20 h to stabilize the color of the PANI layer. In this study, the specimen surface on which the Ni or Ni/PANI layer was formed is referred to as the hydrogen detection side, while the opposite surface is designated as the hydrogen entry side.

For potentiodynamic polarization and galvanic current measurements, a bare steel and Zn plate (99.5%) were used. The specimens were mirror-polished with SiC paper and a diamond paste (6 μm, 1 μm). All experiments were conducted in ambient air at room temperature (*ca.* 293 K).

## 2.2 Electrochemical hydrogen permeation tests

Electrochemical hydrogen permeation tests were performed using a Devanathan–Stachurski cell [44]. The hydrogen detection side, where the Ni layer was formed, was immersed in a 1 M NaOH aqueous solution and polarized at a constant potential of 0.1 V vs. Hg/HgO (1 M NaOH) for 24 h to measure the background current. For the counter and reference electrodes, a Pt wire and a Hg/HgO (1 M NaOH) electrode were used, respectively. Subsequently, hydrogen was introduced into the hydrogen entry side either by cathodic polarization in 1 M NaOH solution (cathodic charging test) or by immersion in 0.1 or 0.01 M NaCl aqueous solutions (corrosion test). For the corrosion test, a vertical Devanathan–Stachurski cell was employed to position the specimen horizontally relative to the ground.

### 2.2.1 Cathodic charging test

To analyze hydrogen diffusion behavior in the steel substrate, electrochemical hydrogen permeation test was conducted on the bare steel specimens. Following the measurement of the background current, the hydrogen entry side was polarized at a constant current density of $-10$ A $m^{-2}$ for 5 h. The electrode area on both the hydrogen entry and detection sides was *ca.* 20 $mm^2$. A Pt wire was employed as a counter electrode for hydrogen charging.

### 2.2.2 Corrosion test

Electrochemical hydrogen permeation current density $i_{perm}$ resulting from corrosion in NaCl solutions was measured using both bare and Zn-coated steel specimens. Figure 1a shows a schematic of the hydrogen entry side of the Zn-coated steel. For the Zn-coated steel, a 10 mm × 10 mm electrode area was fabricated at the center of the specimen using epoxy resin. For bare steel, a 2 mm × 10 mm electrode area was similarly fabricated. Following background current measurements, the hydrogen entry side was immersed in either 0.1 M NaCl or 0.01 M NaCl for 24 h. Optical images of the hydrogen entry side were captured every 300 s using a digital camera. The open-circuit potential (OCP) was measured using a Ag/AgCl (0.1 M or 0.01 M NaCl) reference electrode. Unless otherwise stated, all potential values in this study refer to the Ag/AgCl (3.33 M KCl) reference electrode. After the corrosion test, the specimens were rinsed with distilled water and dried using $N_2$ gas.

## 2.3 Hydrogen visualization tests

Hydrogen visualization tests were performed using bare and Zn-coated steels to analyze the hydrogen flux distribution in the specimens immersed in NaCl solutions. The condition of the hydrogen entry side was identical to that described in Section 2.2.2.

Figure 1b shows a schematic of the hydrogen detection side, where the dashed red square indicates the position of the electrode area on the hydrogen entry side. The sensor initially appeared blue before the test. Figure 1c illustrates the experimental setup for the hydrogen visualization test. The specimen was set in an acrylic cell and placed horizontally relative to the ground. Optical images of the hydrogen detection side were captured every 300 s using an inverted optical microscope equipped with ×1 objective lens. After the hydrogen visualization test, the hydrogen entry side was rinsed with distilled water and dried with $N_2$ gas. The specimen surface was then analyzed using a micro-focused X-ray fluorescence spectrometer (μXRF). Elemental mapping of Fe was performed under vacuum conditions at an accelerating voltage of 50 kV and an irradiation current of *ca.* 1 mA. A Rh tube was used as the X-ray source, with a spot size of 10 μm. Additionally, surface profiles of the hydrogen entry side of specimens were examined using a one-shot 3D measuring microscope.

**2.4 Potentiodynamic polarization measurements**

The potentiodynamic polarization curves were measured for both the bare steel and Zn plate. After mirror-polishing, a 10 mm × 10 mm electrode area was fabricated on the specimen surface using epoxy resin. The prepared specimens were then stored in air for 24 h. The electrode area was immersed in either 0.1 or 0.01 M NaCl solution, and the OCP was measured for 180 s. The potentiodynamic polarization measurements were subsequent conducted from the OCP at a scanning rate of 23 mV $min^{-1}$. A Pt sheet and a Ag/AgCl (3.33 M KCl) reference electrode were utilized as the counter and reference electrodes, respectively.

### 2.5 Measurements of galvanic current and hydrogen permeation current of bare steel coupled to Zn plate

To analyze the correlation between Zn dissolution and hydrogen entry into bare steel, the galvanic current and hydrogen permeation current were measured simultaneously for bare steel coupled to the Zn plate. After mirror-polishing, an electrode area of 2 mm × 10 mm for the bare steel and 8 mm × 10 mm for the Zn plate were prepared using epoxy resin. Subsequently, the specimens were stored in air for 24 h. The other side of the bare steel was coated with a Ni layer following the procedure described in Section 2.1, and the background current was measured for 24 h under the conditions described in Section 2.2. The specimens were then immersed in either 0.1 or 0.01 M NaCl solution. The galvanic current between the bare steel and the Zn plate was recorded using a zero-resistance ammeter, while the hydrogen permeation current was measured simultaneously.

During the test, the bare steel and Zn plate were positioned facing each other at a distance of approximately 10 mm. Both the galvanic current and hydrogen permeation current were converted to the current density using the electrode area of the hydrogen entry side of the bare steel (*ca.* 20 $mm^2$).

### 2.6 FEM simulation of potential distribution on Zn-coated steel in NaCl

FEM simulations were conducted using COMSOL Multiphysics version 6.2 employing a secondary current distribution model to calculate the potential distribution on Zn-coated steel in 0.1 or 0.01 M NaCl. The model incorporates both activation overpotential and ionic transport of charged ions in the solution, applying Ohm's law in combination with charge balance principles. The correlation between charge transfer and overpotential was established based on potentiodynamic polarization curves. The polarization curves of the steel substrate and Zn plate in the NaCl solutions were obtained through the procedure described in Section 2.4. Simulations were performed to evaluate the potential distribution on the Zn-coated steel in the NaCl solutions. The length of both Zn coating and steel substrate were set as 5 mm (see Supplementary Figure 2 for the geometric configuration). Notably, the FEM simulations in this study did not consider the transport (diffusion, migration) of charged and uncharged species. Additionally, effects such as pH changes and the deposition of corrosion products were not incorporated. The potential distribution was calculated based on Ohm's law, and charge conservation within the domain is described as follows:

$$\nabla i = 0 \quad (1)$$

Here, $i$ denotes the current density vector in the electrolyte. Furthermore, the current density is defined as follows:

$$i = -\sigma \nabla \Phi \quad (2)$$

$\sigma$ denotes the conductivity of the electrolyte and $\Phi$ indicates the potential of the electrode. The anodic dissolution of Zn is considered at the Zn coating interface, defined as the deforming boundary:

$$Zn \rightarrow Zn^{2+} + 2e^{-} \quad (3)$$

The remaining boundaries were defined as non-deforming. The steel substrate was designated as a cathodic reaction site. The parameters used in the FEM simulation are listed in Table II, and further details are provided in the Supplementary Information.

### 2.7 Measurement of potential distribution on Zn-coated steel in NaCl

To analyze the correlation between potential distribution and hydrogen entry sites in NaCl solutions, the potential distribution was measured using a SKP [45–47]. An electrode area of 10 mm × 10 mm was fabricated at the center of the Zn-coated steel using epoxy resin. One half of the electrode area (10 mm × 5 mm) was coated with Zn. On the other side of the specimen, a Ni/PANI layer was fabricated following the procedure described in Section 2.1. To prevent galvanic corrosion of the PANI and Ni layers under high humidity conditions, this layer was sealed with resin and a thin glass sheet [43]. The experimental setup is illustrated in Figure 2. The specimen was placed on the SKP stage and electrically connected to it via a Ni wire. A 200 μL droplet of either 0.1 or 0.01 M NaCl solution was placed on the electrode area. The SKP chamber was maintained at approximately 95% relative humidity (RH) by purging a mixture of humidified and dry air through external humidification and dehumidification system. A stainless-steel probe with a diameter of 500 μm was used for the potential measurement. The scan was performed with a step size of 254 μm and a step number 40 × 40, respectively. Prior to each SKP scan, optical images of both the hydrogen entry and detection sides were captured using digital cameras.

## 3. Results and discussion

### 3.1 Hydrogen permeation behavior in the steel substrate

To analyze the hydrogen permeation behavior in the steel substrate, electrochemical hydrogen permeation tests were conducted using bare steel specimens. Figure 3a presents the time variation of $i_{\mathrm{perm}}$ of the bare steel, and Figure 3b shows an enlarged view of Figure 3a. At 70 s, $i_{\mathrm{perm}}$ started to increase and became almost stable after 800 s. The hydrogen diffusion coefficient $D_{\mathrm{H}}$ of the steel substrate was calculated using the following equation:

$$D_{\mathrm{H}} = \frac{L^2}{2t_{\mathrm{L}}} \tag{4}$$

Here, $t_{\mathrm{L}}$ denotes the intersection of bottom axis and the extrapolation line of the integrated value of $i_{\mathrm{perm}}$. Based on the results of the electrochemical hydrogen permeation test, $t_{\mathrm{L}}$ was calculated to be 130 s. The specimen thickness $L$ was 220 μm. Then, the hydrogen diffusion coefficient of the steel substrate was determined to be $1.86 \times 10^{-10}$ $\mathrm{m^2\ s^{-1}}$.

### 3.2 Hydrogen entry behavior into Zn-coated steel in 0.1 M NaCl

#### 3.2.1 Hydrogen permeation current and corrosion of Zn-coated steel

To examine the effect of Zn coating on hydrogen entry, $i_{\mathrm{perm}}$ of the bare and Zn-coated steels immersed in the 0.1 M NaCl solution was measured for 24 h using the Devanathan–Stachurski cell. Figure 4a shows the optical images of the bare steel during the corrosion test. The white dotted curve indicates the electrode area. Corrosion initiated within 1 h and progressed over time, generating corrosion products. However, $i_{perm}$ was not detected during the test. This result suggests that the amount of hydrogen atoms entering the bare steel was so small that $i_{perm}$ was much smaller than the background current, which was *ca.* 20 nA cm$^{-2}$. Figure 4b displays the optical images of the Zn-coated steel during the corrosion test. A visible color change in the Zn coating was observed after 3 h, presumably due to the dissolution of Zn coating. Figure 4c presents the time variations in OCP, $i_{perm}$, and the color-changed area of the Zn coating. The OCP of the Zn-coated steel decreased to *ca.* −1 V immediately after immersion and then stabilized. For reference, the OCP of the bare steel during the same period (from Figure 4a) is shown as a black dotted curve; it reached −0.62 V at 24 h. The Zn coating reduced the OCP by approximately 0.38 V. Since the electrode potential for Zn dissolution is −1.14 V (concentrations of soluble species: $1.0 \times 10^{-6}$ mol kg$^{-1}$) [48], the observed decrease in OCP of the Zn-coated steel can be attributed to Zn dissolution. Notably, $i_{perm}$ was detected during the corrosion test of the Zn-coated steel and increased over time. These results confirm that the electrode potential of the steel substrate decreased below the electrode potential of the HER (−0.53 V at pH 5.5), thereby enhancing both HER and hydrogen entry into the steel substrate [27, 28].

In addition to the increase in $i_{perm}$, the color-changed area of the Zn coating also increased over time. It is considered that the Zn was dissolved, and steel substrate was exposed at the color-changed areas. The observation suggests that hydrogen entry occurs predominantly at the steel substrate exposed to the solution and that the increase in the exposed area contributed to the rise in $i_{perm}$. Figure 4d shows an enlarged view of the time variations of $i_{perm}$ and the color-changed area of the Zn coating, as shown in Figure 4c. After 2 h, the color change of the Zn coating commenced, and $i_{perm}$ started to increase, indicating that $i_{perm}$ rose as Zn coating dissolution progressed. At 24 h, the $i_{perm}$ reached 55 mA $m^{-2}$, which was more than 10 times higher than its value at 2 h (3.7 mA $m^{-2}$). This substantial increase is likely attributed not only to the growth of the exposed area of the steel substrate but also to changes in the solution chemistry and the surface condition of the steel substrate over time.

**3.2.2. Hydrogen entry site of Zn-coated and bare steels in 0.1 M NaCl**

The hydrogen entry into Zn-coated steel was observed to accelerate as the Zn coating dissolved. However, the hydrogen entry site cannot be determined via the electrochemical hydrogen permeation tests. Therefore, the hydrogen entry behavior in 0.1 M NaCl was visualized using a hydrogenochromic sensor. Figures 5a and 5b show optical images of the hydrogen entry side of the Zn-coated steel before and after the hydrogen visualization test, respectively. Figures 5c and 5d present optical images of the hydrogen detection side. The color of the PANI layer changed from blue to yellow due to the reaction with hydrogen atoms that permeated through the Zn-coated steel. Figure 5e depicts the elemental map of Fe, corresponding to the region shown in Figure 5b, obtained using μXRF. In addition to the slit, regions of the steel substrate previously covered by the Zn coating were exposed due to the dissolution of the Zn coating. The hydrogen entry site—identified as the color-changed area of the PANI layer in Figure 5d—appears to correspond to the exposed areas of the steel substrate. To quantitatively analyze the color change in the PANI layer, image analyses were conducted. The red ($R$), green ($G$), and blue ($B$) values of each pixel in Figures 5c and 5d were extracted, and the brightness $Y$ was calculated using the following equation:

$$Y = 0.299R + 0.587G + 0.114B \quad (5)$$

Additionally, the difference in brightness $\Delta Y$ was determined using:

$$\Delta Y = Y_t - Y_0 \quad (6)$$

Here, $Y_0$ indicates the brightness before the hydrogen visualization test, and $Y_t$ represents the brightness at time $t$. While the color of the PANI layer gradually returns to the original color due to the oxidation reaction with $O_2$ in air [38, 43], previous studies have reported that $\Delta Y$ increases in proportion to the logarithm of the total hydrogen flux [38, 39]. Because the color recovery of the PANI layer due to $O_2$ is significantly slower than the color change caused by the hydrogen atoms, hydrogen distribution can be analyzed based on $\Delta Y$ values [49]. If the PANI layer is fully reduced, the increase in the $\Delta Y$ value stops, and the upper limit of a PANI-based hydrogenochromic sensor was reported to be more than 100 C m$^{-2}$ [43]. Figure 5f presents a contour map of the $\Delta Y$ values calculated from the optical images shown in Figures 5c and 5d. The hydrogen entry site corresponded closely to the regions with high Fe concentration (Figure 5e), indicating exposure of the steel substrate. The hydrogen diffusion coefficient of Zn, which is approximately $1.0 \times 10^{-15}$ m$^2$ s$^{-1}$ [50], is significantly slower than that of the steel substrate. The hydrogen flux $J$ (mol m$^{-2}$ s$^{-1}$) in a metal can be calculated using the following equation:

$$J = \frac{C_0 D_\mathrm{H}}{L} \qquad (7)$$

Here, $C_0$ indicates the surface hydrogen concentration. Assuming the Zn coating and steel substrate have the same $C_0$, the flux of the permeated hydrogen through the Zn coating (thickness: *ca*. 2.8 μm) is approximately 0.04% of the flux through the steel substrate (thickness: *ca*. 220 μm). Thus, it is suggested that hydrogen was not detected in the areas where the Zn coating was intact due to the significantly small hydrogen flux in the coating. These results confirm hydrogen entered the Zn-coated steel primarily through the slit and other areas where the steel substrate was exposed to the solution due to the dissolution of Zn coating.

To evaluate the effect of the Zn coating on hydrogen entry behavior, hydrogen visualization tests were also performed using bare steel specimens. Figure 6a shows an optical image of the hydrogen entry side of the bare steel after immersion in 0.1 M NaCl for 24 h, and Figure 6b displays the corresponding $\Delta Y$ contour map of the hydrogen detection side. Hydrogen entry was observed to occur locally within the rust-formed surface; however, the hydrogen entry sites were smaller than the rust-formed area. This observation is consistent with findings from our previous study, which reported that hydrogen entry behavior into an Fe sheet in 3wt% NaCl solution [39]: hydrogen entry does not occur on the non-corroded Fe surface, but rather through the dissolved Fe surface under the rust-formed area. Following the hydrogen visualization test, surface roughness measurements were performed. Figure 6c shows the enlarged optical image of the corroded area from Figure 6a, captured after the hydrogen visualization test. Figure 6d shows the surface roughness profile of the area shown in Figure 6c, revealing two pits with depths of *ca.* 120 μm. Figure 6e displays the $\Delta Y$ contour map of the hydrogen detection side, corresponding to the region shown in Figures 6c and 6d. An increase in $\Delta Y$ was observed at the pit locations. Figure 6f illustrates the line profiles of $\Delta Y$ and surface depth along the line A−B, as indicated in Figures 6d and 6e. High hydrogen flux was detected at the pits and in the adjacent areas approximately 200−300 μm from the pits, consistent with the previous report [24]. The $\Delta Y$ peaks corresponded to the pit centers, confirming that in the case of bare steel, localized decreases in pH and potential occurred within the pits, thereby promoting hydrogen entry at these locations [39].

It was confirmed that the Zn coating served as the primary anodic reaction site, while the HER and hydrogen entry occurred on the steel substrate exposed to the NaCl solution. Although corrosion of the steel substrate in the Zn-coated steel was suppressed, hydrogen entry was significantly accelerated. In contrast, for the bare steel, corrosion progressed, resulting in the formation of large pits. Local hydrogen entry occurred due to reductions in potential and pH within the pits. However, the maximum hydrogen flux observed in the bare steel was substantially lower than that in the Zn-coated steel.

### 3.2.3 Acceleration of hydrogen entry due to the dissolution of neighboring Zn coating

The hydrogen entry site of Zn-coated steel was identified as the bare steel surface exposed to the NaCl solution. However, the detailed behavior of hydrogen entry remains unclear—specifically, whether a gradient of hydrogen flux exists or whether hydrogen entry proceeds uniformly across the steel substrate. Figure 7 shows optical images of the hydrogen entry and detection sides during the hydrogen visualization test shown in Figure 5. After 600 s, local hydrogen entry was detected, as evidenced the color change in the PANI layer. This local hydrogen entry occurred on the steel substrate near the boundary between the steel substrate and the Zn coating (hereafter referred to as the steel/Zn boundary). Over time, the hydrogen entry site expanded within the slit, and hydrogen entry was observed throughout the slit after 1 h. Subsequently, the dissolution of Zn coating near the steel/Zn boundary led to exposure of the underlying steel substrate. This newly exposed area exhibited accelerated hydrogen entry. These observations suggest that Zn coating near the steel/Zn boundary dissolves in the initial stages of corrosion, resulting in local hydrogen entry in the neighboring steel substrate. The optical images shown in Figure 7 were converted to a video file (see Supplementary Video 1).

At 2 h, there was a distribution in hydrogen flux, although the hydrogen entry had proceeded throughout the slit. This implies that by the time $i_{perm}$ stabilized at 2 h (as shown in Figure 4), hydrogen entry occurred uniformly throughout the slit. The surface hydrogen concentration $C_0$ at 2 h was calculated using the equations (7) and (8).

$$J = \frac{i}{F} \quad (8)$$

Here, $F$ is the Faraday constant (C $mol^{-1}$), and $i$ is the $i_{perm}$ of the Zn-coated steel at 2 h (3.7 mA $m^{-2}$), as shown in Figure 4d. Assuming that hydrogen entry occurred throughout the slit at 2 h, the average $C_0$ value on the slit was calculated to be *ca.* 5.8 wppb (ppb by weight). At 24 h, hydrogen entry was detected throughout the area where steel substrate was exposed, as shown in Figure 7. There was a slight variation in hydrogen distribution within the hydrogen entry site: the color change in the slit was slightly less pronounced than that in the newly exposed steel substrate, but overall, the hydrogen entry was almost uniform. Then, the $C_0$ was calculated under the assumption that $i_{perm}$ at 24 h (from Figure 4c) was uniformly distributed across the exposed steel substrate—comprising both the slit and color-changed region (totaling 52 $mm^2$). $C_0$ on the steel substrate at 24 h was calculated to be *ca.* 0.03 wppm. In comparison to the value after 2 h, $C_0$ increased significantly by 24 h, with an increasing rate of *ca.* 1.4 wppb $h^{-1}$ (see Supplementary Figure 3 for the time variation of the $C_0$ value). It is suggested that $C_0$ is influenced not only by the expansion of the hydrogen entry site, but also changes in solution pH and the surface condition of the steel substrate.

To further analyze the hydrogen entry behavior in detail, image analyses were conducted. Figure 8 presents the time variation of the average $\Delta Y$ values in Areas 1−4, as indicated by the red squares in Figure 7. Area 4 corresponds to the region where the color of the Zn coating rarely changed, and no hydrogen entry was detected. Conversely, the color change of the Zn coating was observed in Area 3, and the $\Delta Y$ value rapidly increased after 19 h. These observations indicate that hydrogen entry does not occur while the Zn coating is intact, but begins only after the steel substrate becomes exposed to the solution. Areas 1 and 2 correspond to the slit area. In Area 1, the $\Delta Y$ value drastically increased at the beginning of the test, whereas in Area 2, the increase was more gradual. However, after 21 h, a sharp rise in $\Delta Y$ value was observed in Area 2—from 90 to 110. Eventually, the $\Delta Y$ values in both Areas 1 and 2 became almost the same, suggesting that hydrogen entry did not proceed uniformly on the steel substrate. This supports the presence of a hydrogen flux distribution within the steel substrate.

As was shown in Figure 7, hydrogen entry initially occurs locally on the steel substrate near the steel/Zn boundary. The local hydrogen entry was accompanied by the dissolution of the neighboring Zn coating, implying that the dissolution process plays a critical role in facilitating hydrogen entry. Figures 9a and 9b present enlarged views of Areas 1 and 2, respectively. Figures 9d and 9e show the time variation of the $\Delta Y$ values in these areas, with the white dots indicating the time points at which the optical micrographs in Figures 9a and 9b were captured. Up to 1 h, color change was not confirmed in the Zn coating. In Area 1, the $\Delta Y$ value stabilized at 96 after 1.3 h and subsequently increased again at 3 h. This second increase coincided with the onset of color change in the Zn coating on the right side of Area 1. This suggests that the enhanced hydrogen entry (as reflected in the $\Delta Y$ value increased between 3 to 5 h) was attributable to the acceleration of HER in Area 1 due to the dissolution of Zn coating on the right side of Area 1. From 5 to 15 h, the $\Delta Y$ value remained relatively stable. Given that the $\Delta Y$ value of the PANI layer gradually decreases over time due to the oxidation by $O_2$ in air, when hydrogen charging is stopped [38, 43], a stable $\Delta Y$ value from 5 to 15 h indicates that the hydrogen flux remained consistent during this period. The $\Delta Y$ value increased again between 15 to 20 h, corresponding to the onset of Zn coating dissolution on the left side of Area 1, as indicated by the blue arrow. After 20 h, a visible color change in the Zn coating was confirmed in the region marked by the blue curve. These findings suggest that the increase in $\Delta Y$ between 15 and 20 h was caused by the dissolution of Zn coating on the left side of Area 1. Overall, the results confirm that the dissolution of Zn coating promotes HER on the adjacent steel substrate, thereby further enhancing local hydrogen entry.

As was shown in Figures 9b and 9e, the $\Delta Y$ value in Area 2 reached approximately 65 at 5 h, even though no color change was observed in the adjacent Zn coating. The $\Delta Y$ value continued to increase with time and stabilized after 10 h. Compared to Area 1, the $\Delta Y$ value in Area 2 remained relatively low, likely due to the greater distance between Area 2 and the dissolving Zn coating, which was shorter in the case of Area 1. From 10 to 21 h, little to no color change was observed in the Zn coating adjacent to Area 2, indicating that HER activity remained limited during this period. However, at 21.5 h, Zn coating dissolution commenced on the left side of Area 2, as indicated by the blue arrow, and this was followed by the increase in the $\Delta Y$ value. These results confirm that hydrogen entry into the steel substrate is accelerated by the dissolution of adjacent Zn coating. The dissolution process of Zn continued, and after 24 h, the underlying steel substrate was exposed in the region marked by the blue curve. Figures 9c and 9f present the optical images and time variation of the $\Delta Y$ value in Area 3. In this region, neither color change nor evidence of hydrogen entry was confirmed until 18 h. After this point, the color of the Zn coating started to change, and subsequently, the $\Delta Y$ value drastically increased after 19 h. These results confirm that hydrogen entry is minimal while the Zn coating remains intact on the steel substrate, but it is significantly accelerated as the steel substrate becomes exposed to the solution.

A non-uniform hydrogen flux distribution was observed within the hydrogen entry site, where the steel substrate was exposed to the solution. The hydrogen flux was notably higher in the steel substrate adjacent to the dissolving Zn coating. In addition, no hydrogen entry occurred in areas where the Zn coating remained intact.

### 3.3 Effect of $Cl^-$ concentration on hydrogen entry behavior into Zn-coated steels

Although $Cl^-$ concentration is known to influence hydrogen entry into steels [25, 51, 52], its specific role in Zn-coated steels remains unclear. In this section, hydrogen entry behavior in Zn-coated steels was investigated using a NaCl solution with lower $Cl^-$ concentration than that used in Section 3.2, in order to investigate the effect of $Cl^-$ concentration on hydrogen uptake. Figure 10a shows the optical images of the Zn-coated steel before (left) and after (right) the corrosion test in 0.01 M NaCl for 24 h. The color in the Zn coating indicates corrosion. For comparison, Figure 10b presents optical images of the Zn-coated steel before and after a similar corrosion test in 0.1 M NaCl for 24 h, as shown previously in Figure 4b. No significant difference in surface appearance was observed between specimens immersed in 0.01 and 0.1 M NaCl. The color-changed areas of the Zn coating after testing were *ca.* 39 $mm^2$ and *ca.* 32 $mm^2$ for the 0.01 and 0.1 M

NaCl solutions, respectively. These results suggest that decreasing $Cl^-$ concentration from 0.1 to 0.01 M had minimal impact on the macroscopic dissolution behavior of Zn coating.

Figure 11a shows the time variation of the OCP and $i_{perm}$ of the Zn-coated steel during the corrosion test in 0.01 M NaCl. For reference, the corresponding results of the Zn-coated steel in 0.1 M NaCl (Figure 4c) are indicated by the dotted curves. The OCP in 0.01 M NaCl was *ca.* 0.1 V higher than that in 0.1 M NaCl. Regardless of $Cl^-$ concentration, $i_{perm}$ increased in both solutions; however, the magnitude was higher in 0.1 M NaCl than in 0.01 M NaCl. Figure 11b presents the time variation in the color-changed area of the Zn coating. The growth rate of this area was found to be independent of $Cl^-$ concentration, suggesting that the dissolution rate of the Zn coating was nearly identical in both solutions. After 10 h, the growth rate of this area was approximately 1.4 $mm^2$ $h^{-1}$ (see Supplementary Figure 4 for detail). Figure 11c presents an enlarged view of the time variations of $i_{perm}$ and the color-changed area, as shown in Figures 11a and 11b. In 0.01 M NaCl, $i_{perm}$ was first detected after 0.6 h (21.6 ks), approximately 720 s later than in 0.1 M NaCl. In both 0.01 and 0.1 M NaCl, $i_{perm}$ reached a stable value before any visible color change was observed in the Zn coating. When $i_{perm}$ stabilized at 2 h, the $i_{perm}$ in 0.1 M NaCl was 3.7 mA $m^{-2}$, while it was 0.7 mA $m^{-2}$ in 0.01 M NaCl. This suggests that hydrogen entry was suppressed in 0.01 M NaCl compared with that in 0.1

M NaCl, although there is no significant difference in the overall corrosion behavior of the Zn coating.

To elucidate the effect of $Cl^-$ concentration on the hydrogen entry site and hydrogen flux distribution, a hydrogen visualization test was conducted in 0.01 M NaCl. Figure 12 presents optical images of the hydrogen entry and detection sides during the test. In 0.1 M NaCl, hydrogen entry was detected at 600 s, whereas in 0.01 M NaCl, it was first observed at approximately 0.5 h. The initial hydrogen entry site was the steel substrate located near the steel/Zn boundary of the slit, consistent with the results obtained in 0.1 M NaCl. Even after 5 h, there was a clear positional dependence of the hydrogen flux within the slit: regions of the steel substrate adjacent to the dissolving Zn coating exhibited higher hydrogen flux, which decreased with increasing distance from the dissolving Zn coating. The hydrogen entry site expanded with time, with additional hydrogen entry also detected in newly exposed areas of steel substrate due to the continued dissolution of the Zn coating. The optical images shown in Figure 12 were converted to a video file (see Supplementary Video 2)

Figures 13a and 13b show the optical image of the corrosion side and the corresponding Fe element map, respectively, obtained after the hydrogen visualization test shown in Figure 12. Figure 13c depicts the optical image of the hydrogen detection side. The hydrogen entry site roughly corresponded to the area where the steel substrate was exposed, indicating that the overall size of the hydrogen entry site was independent of the $Cl^-$ concentration. However, the hydrogen flux distribution was non-uniform in 0.01 M NaCl compared to that observed in 0.1 M NaCl.

To analyze the local hydrogen entry behavior in detail, the average $\Delta Y$ values were calculated in Areas 1−3, as indicated by the red squares in Figure 13c. Area 1 corresponds to the area within the slit where hydrogen flux was the highest. The $\Delta Y$ value in this area began to increase at 1200 s and reached a stable value of 102 at 2 h. Subsequently, between 6 and 10 h, the $\Delta Y$ value further increased to 115, attributed to the dissolution of the Zn coating on the left side of Area 1. In Area 2, the $\Delta Y$ value gradually increased after 4 h, and stabilized after 15 h. The hydrogen entry observed in Area 2 was presumably due to the gradual expansion of the dissolving Zn coating, which enhanced HER in that region over time. In contrast, no hydrogen entry was detected in Area 3, as the steel substrate remained unexposed.

Hydrogen visualization tests revealed distinct differences in the hydrogen flux distribution between 0.1 and 0.01 M NaCl. Then, to quantitatively assess these differences, contour maps of the $\Delta Y$ values on the hydrogen detection sides were generated. Figures 14a and 14b present the $\Delta Y$ contour maps obtained from the hydrogen visualization tests with 0.1 (Figure 7) and 0.01 M NaCl (Figure 12), respectively. In 0.1 M NaCl, hydrogen entry was detected in approximately half of the slit by 0.5 h, following an initial increase in the $\Delta Y$ value at 600 s. In contrast, in 0.01 M NaCl, hydrogen entry was observed only near the steel/Zn boundary at 1 h. The final sizes of the hydrogen entry sites in 0.1 and 0.01 M NaCl were 87 and 77 mm$^2$, respectively, showing no significant dependence on $Cl^-$ concentration. However, a clear difference in the distribution of the $\Delta Y$ values (reflecting hydrogen flux) was observed between the two NaCl concentrations. In particular, the hydrogen flux distribution within the slit marked variation: in 0.1 M NaCl, the $\Delta Y$ values were predominantly high, with a maximum value of 129. In contrast, in 0.01 M NaCl, significant hydrogen entry occurred only locally—within *ca.* 1-mm area of the dissolving Zn coating—with a maximum value of 127. In other regions, the $\Delta Y$ values ranged from *ca.* 30 to 50. These results indicate that the maximum hydrogen flux was not largely dependent by $Cl^-$ concentration. However, the hydrogen flux was non-uniform in 0.01 M NaCl compared with that in 0.1 M NaCl. Given that the size of the hydrogen entry site was independent of $Cl^-$ concentration, the observed difference in $i_{perm}$ between 0.1 and 0.01 M (as shown in Figure 11a) is attributed to the variation in hydrogen flux distribution in the NaCl solutions.

## 3.4 Critical factor for hydrogen flux distribution: galvanic current vs. potential distribution

### 3.4.1 Effect of $Cl^-$ concentration on galvanic current

As the $Cl^-$ concentration was decreased from 0.1 to 0.01 M, the hydrogen flux distribution significantly changed. This change may be attributed to a decrease in the galvanic current between the steel substrate and the Zn coating. Lower $Cl^-$ concentration could reduce the dissolution rate of the Zn coating, thereby diminishing the galvanic current. Consequently, HER may be confined to areas on the steel substrate located in close proximity—*ca.* 1 mm—to the actively dissolving Zn coating. To investigate the influence of $Cl^-$ concentration on galvanic current, potentiodynamic polarization curves of the bare steel and Zn plate were measured in the NaCl solutions. Figure 15 presents the anodic polarization curves of the Zn plate (solid curve) and the cathodic polarization curves of the bare steel (dotted curve) in 0.1 and 0.01 M NaCl. The cathodic polarization behaviors of the bare steel remained nearly identical in both NaCl solutions. In contrast, the anodic polarization curve of the Zn plate shifted in a more noble direction with decreasing $Cl^-$ concentration. Despite this shift, the current densities at the intersection points of the cathodic and anodic polarization curves in 0.1 and 0.01 M NaCl were 0.26 and 0.27 A $m^{-2}$, respectively. Therefore, these results suggest that the galvanic current between the steel substrate and Zn coating is barely affected by the reduction in $Cl^-$ concentration.

To confirm the effect of $Cl^-$ concentration on the galvanic current between the steel substrate and Zn plate, galvanic current measurements were conducted in the NaCl solutions. To simulate the galvanic reactions of the Zn-coated steel, electrode areas of 8 mm × 10 mm and 2 mm × 10 mm were fabricated on the Zn plate and bare steel, respectively. Figure 16a shows the time variations of the galvanic current density of the bare steel coupled to the Zn plate in 0.1 and 0.01 M NaCl. The galvanic current density is reported as positive when the Zn plate act as the anode. After 5 h, the galvanic current densities in 0.1 and 0.01 M NaCl were 63 and 74 mA $m^{-2}$, respectively. Based on the results shown in Figures 15 and 16a, the galvanic current between the steel substrate and Zn coating was found to be almost the same in both 0.1 and 0.01 M NaCl. Figure 16b and 16c present the time variations of the OCP and $i_{perm}$ of the bare steel coupled to Zn plate shown in Figure 16a. The OCP in 0.1 M NaCl was lower than that in 0.01 M NaCl, and corresponding $i_{perm}$ was higher aligning with those obtained from the electrochemical hydrogen permeation tests (Figure 11a). Figure 17 displays the integrated values of galvanic current density ($C_{gal}$) and $i_{perm}$ ($C_{perm}$) for 0.1 and 0.01 M NaCl (shown in Figures 16a and 16c). While $C_{gal}$ was almost the same between 0.1 and 0.01 M NaCl, $C_{perm}$ in 0.1 M NaCl was approximately four times larger than in 0.01 M NaCl. Therefore, it was concluded that the galvanic current barely changes with the decrease in $Cl^-$ concentration from 0.1 to 0.01 M, and its influence on the decrease in $i_{perm}$ (Figures 11a, 11b, and 16c) is negligible. The observed difference in $C_{perm}$ is likely caused by a shift in the ratio between the ORR and HER for cathodic current in 0.1 and 0.01 M NaCl solutions. This shift is presumed to result from changes in the potential distribution on the steel surface as a function of $Cl^-$ concentration.

### 3.4.2 Potential distribution of Zn-coated steel in NaCl solutions

The electric conductivities of 0.01 and 0.1 M NaCl utilized in this study were 0.11 and 1.0 mS $cm^{-1}$, respectively. When current flows through an electrolyte, lower conductivity results in a greater potential decrease (or increase), leading to the formation of a more pronounced potential gradient along the metal surface [7, 53]. Accordingly, a steeper potential gradient was expected to be generated on the Zn-coated steel in 0.01 M NaCl: a decrease in the potential below the electrode potential of HER would occur only near the actively dissolving Zn coating, while regions farther from the dissolution site would retain higher potential. To investigate this hypothesis, the potential distribution on Zn-coated steel was simulated using FEM. Figure 18a presents the simulated potential distribution on Zn-coated steel in 0.1 and 0.01 M NaCl solutions. The black-shaded region corresponds to the Zn coating. In 0.1 M NaCl, the potential distribution across the steel substrate remained nearly uniform. In contrast, in 0.01 M NaCl, the potential on steel substrate decreased as the position approached the steel/Zn boundary. To further analyze this effect, the potential dependence of current density of HER ($i_H$) was estimated using the cathodic polarization curves shown in Figure 15. The linear increase in current density in the potential range lower than −1 V was attributed to HER, with this region extrapolated as a function of potential (see Supplementary Figure 5 for detail). The distribution of $i_H$ on the steel substrate was calculated using the linear relationship between the potential and $i_H$ in combination with potential distribution shown in Figure 18a. As illustrated in Figure 18b, $i_H$ was constant in 0.1 M NaCl, while a gradient of $i_H$ was confirmed in 0.01 M NaCl. As the position approached the steel/Zn boundary, $i_H$ increased. Tada *et al*. reported that the potential gradient on an Fe surface coupled with a Zn coating is more pronounced in 0.01 M NaCl than that in 0.1 M NaCl [34]. Specifically, they observed a potential gradient exceeding 10 mV $mm^{-1}$ near the steel/Zn boundary in 0.01 M NaCl solution, and the gradient decreased with increasing NaCl concentration. Okada *et al*. experimentally and theoretically investigated the potential distribution on a steel coupled with Zn coating in a NaCl solution [54]. They reported a potential gradient of approximately 10 mV mm$^{-1}$ near the steel/Zn boundary in a 1000 ppm (0.017 M) thin NaCl electrolyte (thickness: 2 mm). It was determined that the potential distribution on Zn-coated steel was changed due to differences in electric conductivity between 0.1 and 0.01 M NaCl solutions, thereby inducing a gradient of $i_H$ across the steel substrate.

The potential gradients in 0.01 M NaCl simulated in this study were lower than those reported in previous studies [34, 36, 54, 55], likely due to exclusion of changes in solution chemistry and electrode surface conditions. To experimentally validate the effect of $Cl^-$ concentration on the potential distribution of Zn-coated steel, SKP measurements were conducted in both NaCl solutions. Figures 19a and 19d show optical images of the hydrogen entry side after 7.2 h of immersion in 0.1 and 0.01 M NaCl, respectively. The solid and dotted curves indicate the electrode area and the steel/Zn boundary, respectively. At 7.2 h, an electrolyte film existed on the electrode area, and thus, the concentration of the NaCl solution was not in equilibrium (see Supplementary Figure 6 for detailed information). Figures 19b and 19e show the potential distributions of Zn-coated steel after 7.2 h of immersion in 0.1 and 0.01 M NaCl. In 0.1 M NaCl, the potential across the entire steel substrate was uniformly low. In 0.01 M NaCl, the potential was low only near the Zn/Fe boundary and gradually increased with distance from the steel/Zn boundary. These results are consistent with the FEM calculations presented in Figure 18 and previous studies [34, 53, 54]. In this study, the area exhibiting decreased potential on the steel substrate was confirmed to be smaller in 0.01 M NaCl than in 0.1 M NaCl. Figures 19c and 19f display optical images of the hydrogen detection side after 7.2 h of immersion in 0.1 and 0.01 M NaCl, respectively. In 0.1 M NaCl, hydrogen entry occurred across the entire steel substrate, whereas in 0.01 M NaCl, it was confined to regions near the steel/Zn boundary. These findings indicate that the low electrical conductivity of 0.01 M NaCl gives rise to a potential distribution on the steel substrate, which in turn leads to hydrogen flux distribution: the hydrogen flux is high in steel substrate *ca.* 1−2 mm from the dissolving Zn coating, while it remains low in more distant areas.

### 3.5 Hydrogen entry behavior into Zn-coated steels in 0.1 M and 0.01 M NaCl

In 0.1 M NaCl, bare steel is corroded, leading to the formation of pits. Hydrogen entry is facilitated within these pits due to local decreases in pH and potential. In such cases, the ORR predominantly occurs outside the corroded regions, while the HER is promoted with in the corroded area due to the acidification induced by the hydrolysis reaction of Fe ions [39]. In contrast, Zn-coating prevents the corrosion of the steel substrate through a sacrificial corrosion protection mechanism: the Zn coating works as the anodic reaction site, while cathodic reactions occur on the underlying steel substrate. Despite this protection, HER is significantly accelerated on the steel—even under neutral or alkaline conditions, resulting in substantial hydrogen entry. This occurs because the electrode potential for Zn dissolution (−1.1 V at a soluble species concentration of $1.0 \times 10^{-6}$ mol kg$^{-1}$) is much lower than that for HER (−0.53 V at pH 5.5).

In the initial stage of corrosion of Zn-coated steel in an NaCl solution, the Zn coating near the steel/Zn boundary begins to dissolve, leading to local hydrogen entry on the steel substrate on adjacent regions of exposed steel substrate. Hydrogen entry into the steel substrate beneath intact Zn coating is minimal; however, it is accelerated rapidly once the steel substrate is exposed to the NaCl solution. Consequently, as the Zn coating dissolves, and hydrogen entry sites expand in parallel with the growing area of exposed steel substrate. Hydrogen entry was significantly accelerated in regions adjacent to dissolving Zn coating, where HER on the steel substrate was promoted by the anodic dissolution of neighboring Zn coating. In 0.1 M NaCl, $C_0$ on the steel substrate was approximately 5.7 wppb during the initial corrosion process, increasing over time to 0.03 wppm after 24 h. The increasing rate of $C_0$ was *ca.* 1.4 wppb h$^{-1}$. Assuming a constant rate, $C_0$ would reach 0.04 wppm after 29 h—equivalent to $C_H$ for AISI4235 steel with 1500 MPa tensile strength [26]. For comparison, AISI4135 steel has been shown to exceed 0.04 wppm in diffusible hydrogen content after 60 cycles of salt spray cyclic corrosion test at 303 K [23]. This indicate that Zn coating may accelerate the hydrogen entry and increase the risk of hydrogen embrittlement. The time-dependent increase in $C_0$ is likely influenced by changes in solution pH and the evolving surface state condition of the steel substrate [36, 55]; however, further investigation is required to elucidate their respective roles in hydrogen entry.

The total amount of hydrogen entry into the Zn-coated steel in 0.01 M NaCl was lower than in 0.1 M NaCl, primarily due to differences in hydrogen flux distribution: In 0.1 M NaCl, the hydrogen flux became relatively uniform and high across the entire steel substrate exposed to the solution after 5 h. In contrast, in 0.01 M NaCl, the hydrogen flux became high only near the dissolving Zn—even after 24 h—and diminished with increasing distance from the actively dissolving Zn coating. Notably, neither the galvanic current nor the overall dissolution behavior of Zn coating showed substantial differences between 0.1 and 0.01 M NaCl solutions. This indicates the altered the hydrogen flux distribution was not attributed to changes in galvanic current. Although, the OCP increased as $Cl^-$ concentration decreased—potentially reducing the driving force for HER—this shift alone cannot account for the significant potential gradient observed in

0.01 M NaCl.

The electric conductivities of 0.01 and 0.1 M NaCl were 0.11 and 1.0 mS cm$^{-1}$, respectively. When current flows through an electrolyte, both current and potential distributions are generated on the metal surface, and these distributions are influenced by the electric conductivity, as indicated by Equation (2). Therefore, when the same galvanic current flows on the electrode area, the smaller the electric conductivity, the larger the gradient of the potential distribution. In 0.1 M NaCl, the potential was almost constant at approximately −1 V in 0.1 M NaCl. However, in 0.01 M NaCl, the potential was lower than the electrode potential of HER only in the vicinity of the dissolving Zn and increased with distance from this region. As a result, while the maximum hydrogen flux was not largely different, the hydrogen flux distribution varied significantly between 0.1 and 0.01 M NaCl, leading to variation in the total hydrogen flux, $C_{perm}$.

In this study, the roles of the galvanic current and potential distribution in hydrogen entry were clarified through hydrogen visualization tests, FEM simulations, and potential measurement using an SKP. While the present study focused on galvanic current and potential distribution, a more comprehensive understanding of the hydrogen entry mechanism into Zn-coated steels will require future studies that explicitly consider changes in solution chemistry [35, 36, 56] and the surface state of the steel substrate [31,

34, 57, 58].

## Conclusions

The hydrogen entry behavior of Zn-coated steel in NaCl solutions was successfully visualized *in situ* using a polyaniline (PANI)-based hydrogenochromic sensor. Electrochemical hydrogen permeation tests and potential measurements using a scanning Kevin probe were conducted to clarify the effects of galvanic current and potential distribution on hydrogen entry. The main conclusions are as follows.

・In 0.1 M NaCl, pits formation occurred on the bare steel surface, and hydrogen entry was accelerated in the pits due to the local decreases in pH and electrode potential.

・The corrosion of the steel substrate in Zn-coated steel was mitigated by the sacrificial protection effect of Zn coating. However, the dissolution of Zn reduced the potential of the steel substrate below the electrode potential for HER (−0.53 V at pH 5.5), thereby accelerating HER and hydrogen entry into the steel substrate.

・Hydrogen entry did not proceed uniformly throughout the steel substrate. Instead, the hydrogen flux was significantly high near the regions adjacent to the dissolving Zn coating.

・No significant differences in dissolution behavior of the Zn coating and the galvanic current were observed between 0.1 and 0.01 M NaCl solutions. However, the total hydrogen flux in 0.01 M NaCl reduced by approximately 75% (with decreasing $Cl^-$ concentration) of that in 0.1 M. This reduction is attributed to changes in the potential distribution caused by the lower electrical conductivity of the NaCl solution at reduced $Cl^-$ concentration. Specifically, in 0.1 M NaCl, the hydrogen flux distribution was a relatively uniform across the exposed steel substrate. In contrast, in 0.01 M NaCl, the hydrogen flux only near the dissolving Zn coating (*ca.* 1−2 mm away) was comparable to that in 0.1 M NaCl due to a local decrease in potential. However, as the distance from the dissolving Zn coating increased, the potential on the steel substrate rose accordingly, reducing the driving force for HER and limiting hydrogen entry.

・In 0.1 M NaCl, $C_0$ on the steel substrate was approximately 5.7 wppb in the initial corrosion process, but it increased over time and increased to 0.03 ppm after 24 h. The increasing rate of $C_0$ was *ca.* 1.4 wppb $h^{-1}$.

・The difference in hydrogen flux distribution between 0.1 and 0.01 M NaCl solutions was not attributed to change in galvanic current. Rather, the predominant factor influencing hydrogen flux distribution was the potential distribution on the steel substrate.

## Data availability

The authors confirm that the data supporting the findings of this study are available in the article and its supplementary materials.

**Acknowledgement**

This work was supported by JSPS KAKENHI Grant Number JP25K01541.

**Contributions**

**Hiroshi Kakinuma**: writing—original draft, conceptualization, investigation, methodology, data curation, formal analysis, funding acquisition, project administration. **Saya Ajito**: writing—review & editing, investigation, data curation, methodology. **Koki Okumura**: writing–review & editing, investigation. **Makoto Akahoshi**: writing—review & editing, investigation, resources. **Yu Takabatake**: writing—review & editing, investigation. **Tomohiko Omura**: writing—review & editing, resources, funding acquisition, project administration. **Motomichi Koyama**: writing—review & editing. **Eiji Akiyama**: writing—review & editing, funding acquisition, project administration.

**Tables and Figures**

Table I. Chemical composition of the steel sheet used in this study (mass%)

| C | Si | Mn | P | S | Fe |
|---|---|---|---|---|---|
| 0.037 | 0.018 | 0.24 | 0.014 | 0.0062 | Bal. |

Table II. Parameters used for FEM simulations

| Parameter | Description | Value |
|---|---|---|
| $d_{Zn}$ | Density of Zn | $7.1\times10^{3}$ kg m$^{-3}$ |
| $M_{Zn}$ | Molar mass of Zn | 65 g mol$^{-1}$ |
| $\sigma_{0.1}$ | Conductivity of 0.1 M NaCl | 1.0 mS cm$^{-1}$ |
| $\sigma_{0.01}$ | Conductivity of 0.01 M NaCl | 0.11 mS cm$^{-1}$ |
| $E_{Fe}$ | Equilibrium potential of Fe | −0.25 V |
| $E_{Zn}$ | Equilibrium potential of Zn | −1 V |

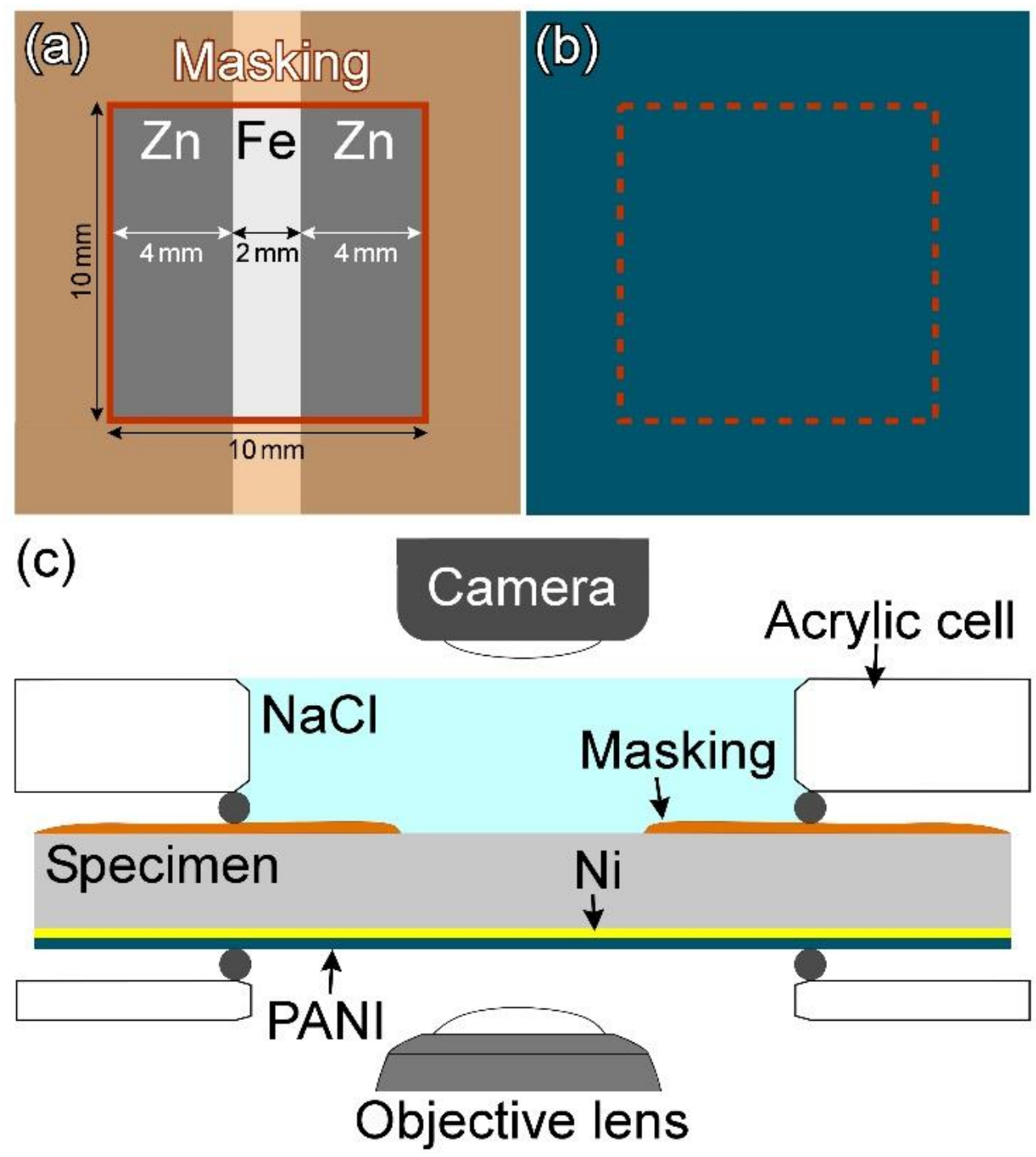

**Figure 1.** Schematics of (a) hydrogen entry and (b) detection sides of the Zn-coated steel. The red square in (a) indicates the electrode area, and the corresponding area is indicated by the dashed red square in (b). (c) Experimental setup for hydrogen visualization test.

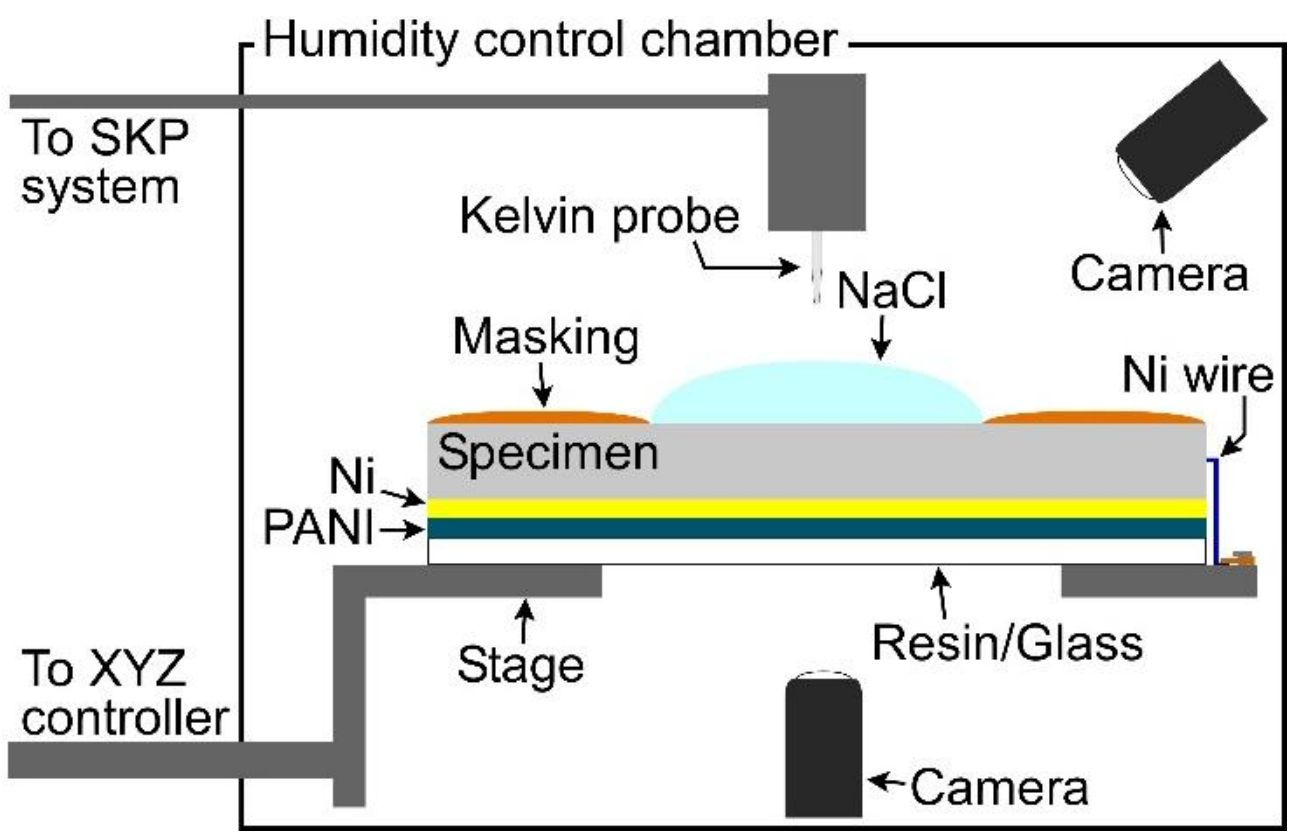

**Figure 2.** Experimental setup for measurement of potential distribution and hydrogen visualization.

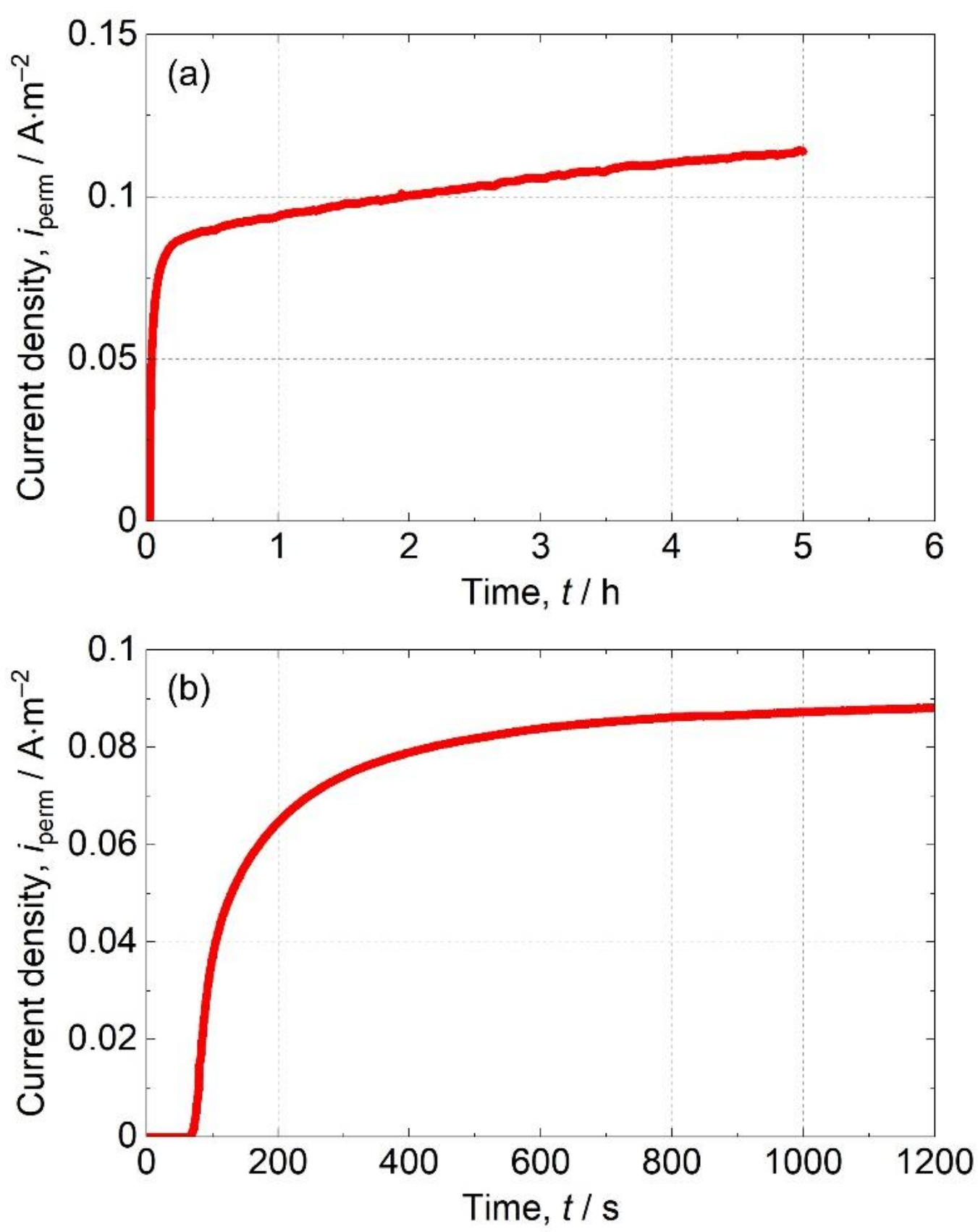

**Figure 3**. (a) Time variation of the $i_{perm}$ of the bare steel and (b) enlarged view of the initial portion of (a).

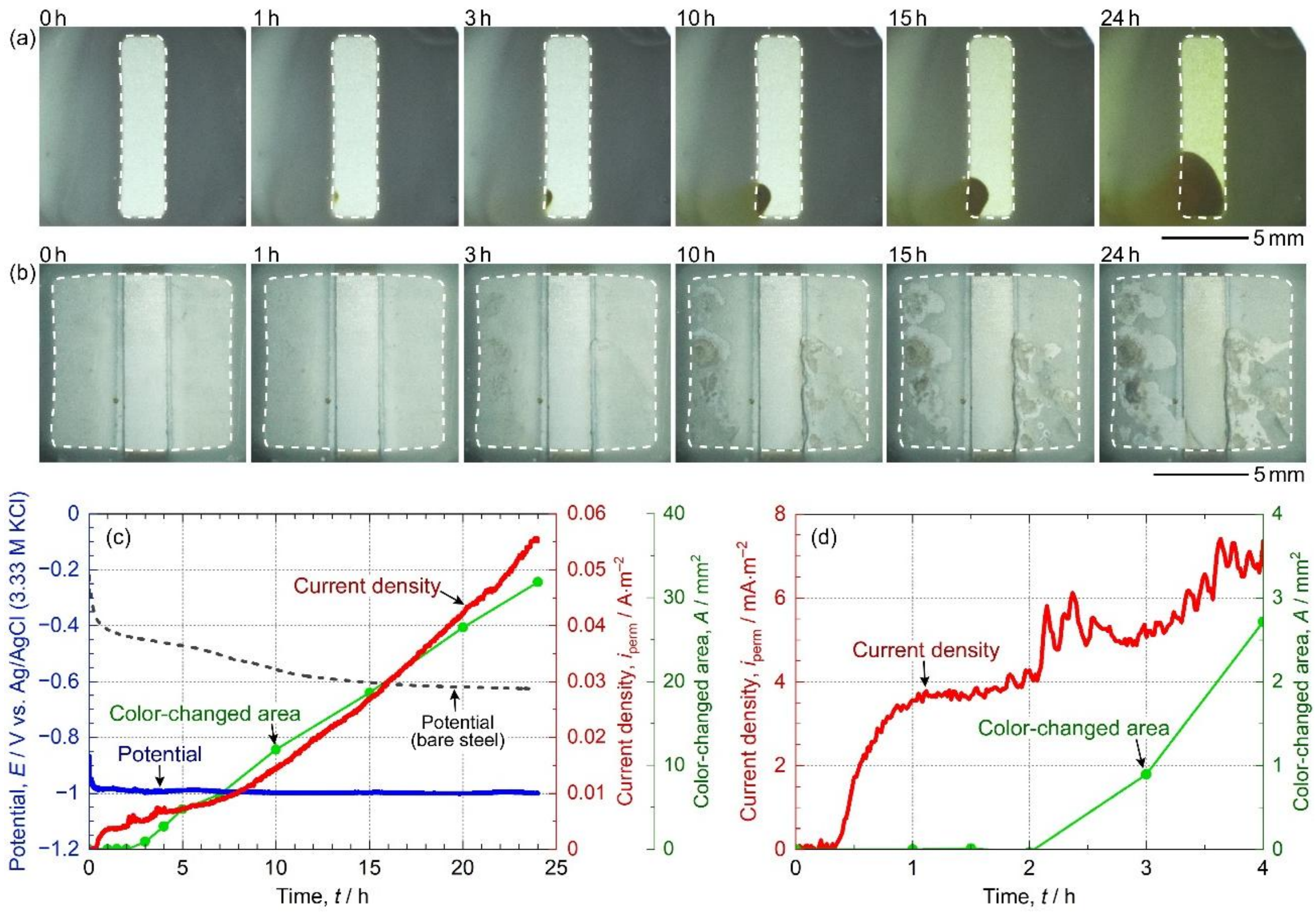

**Figure 4.** Optical images of the hydrogen entry side of (a) the bare and (b) Zn-coated steels during corrosion tests in 0.1 M NaCl. (c) Time variations of OCP (blue), $i_{perm}$ (red), and color-changed area (green) of Zn-coated steel during the corrosion test shown in (b). The OCP of the bare steel is indicated by the black dotted curve for reference. (d) Enlarged view of $i_{perm}$ and color-changed area shown in (c).

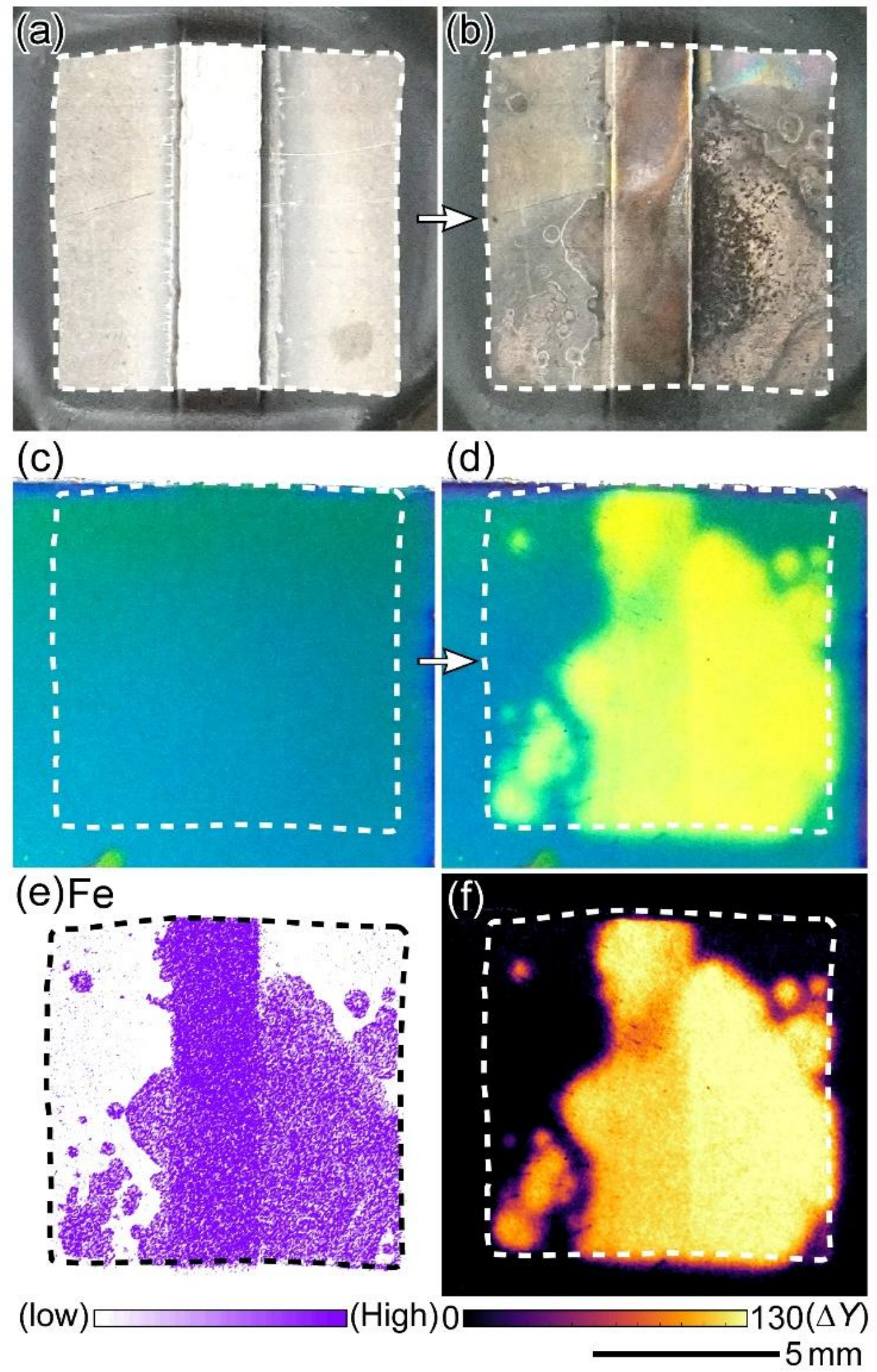

**Figure 5.** Optical images of (a, b) hydrogen entry and (c, d) detection sides before and after the hydrogen visualization test in 0.1 M NaCl. (e) Elemental map of Fe on the hydrogen entry side corresponding to (c). (f) Contour map of $\Delta Y$ on the hydrogen detection side, calculated using (c) and (d).

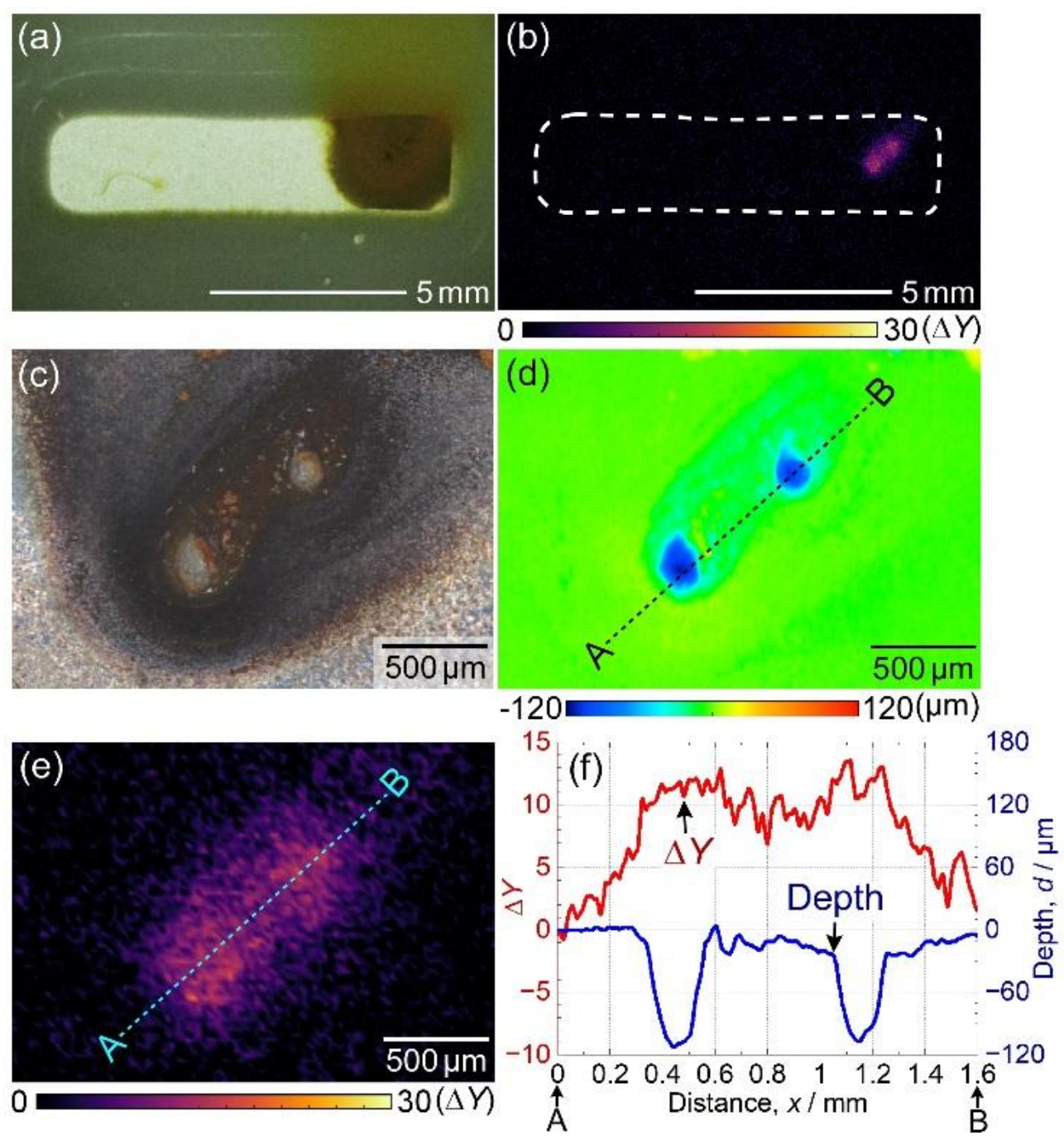

**Figure 6.** (a) Optical image of hydrogen entry side of bare steel and (b) the corresponding $\Delta Y$ contour map of the hydrogen detection side after 24 h during the hydrogen visualization test. The white dotted curves indicate the electrode area on the hydrogen entry side. (c) Optical micrograph of the corroded area after the test shown in (a). (d) Surface roughness profile and (e) $\Delta Y$ contour map of the area corresponding to (c). (f) Line profiles of surface depth and $\Delta Y$ value along the line A−B indicated in (d) and (e).

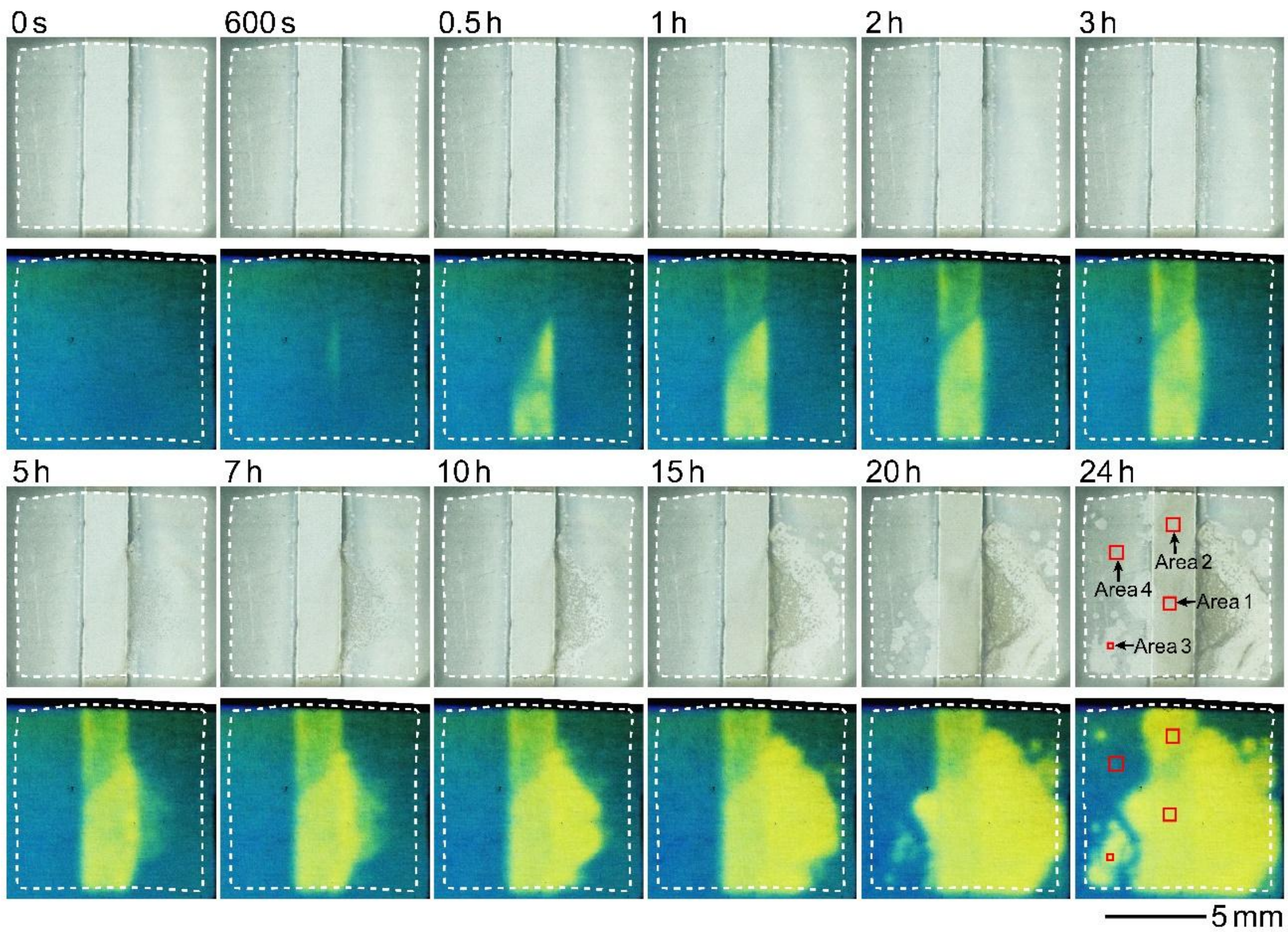

**Figure 7.** Optical images of hydrogen entry (upper row) and detection (lower row) sides during the hydrogen visualization test in 0.1 M NaCl (Figure 5). The white dotted curves indicate the electrode area on the hydrogen entry side.

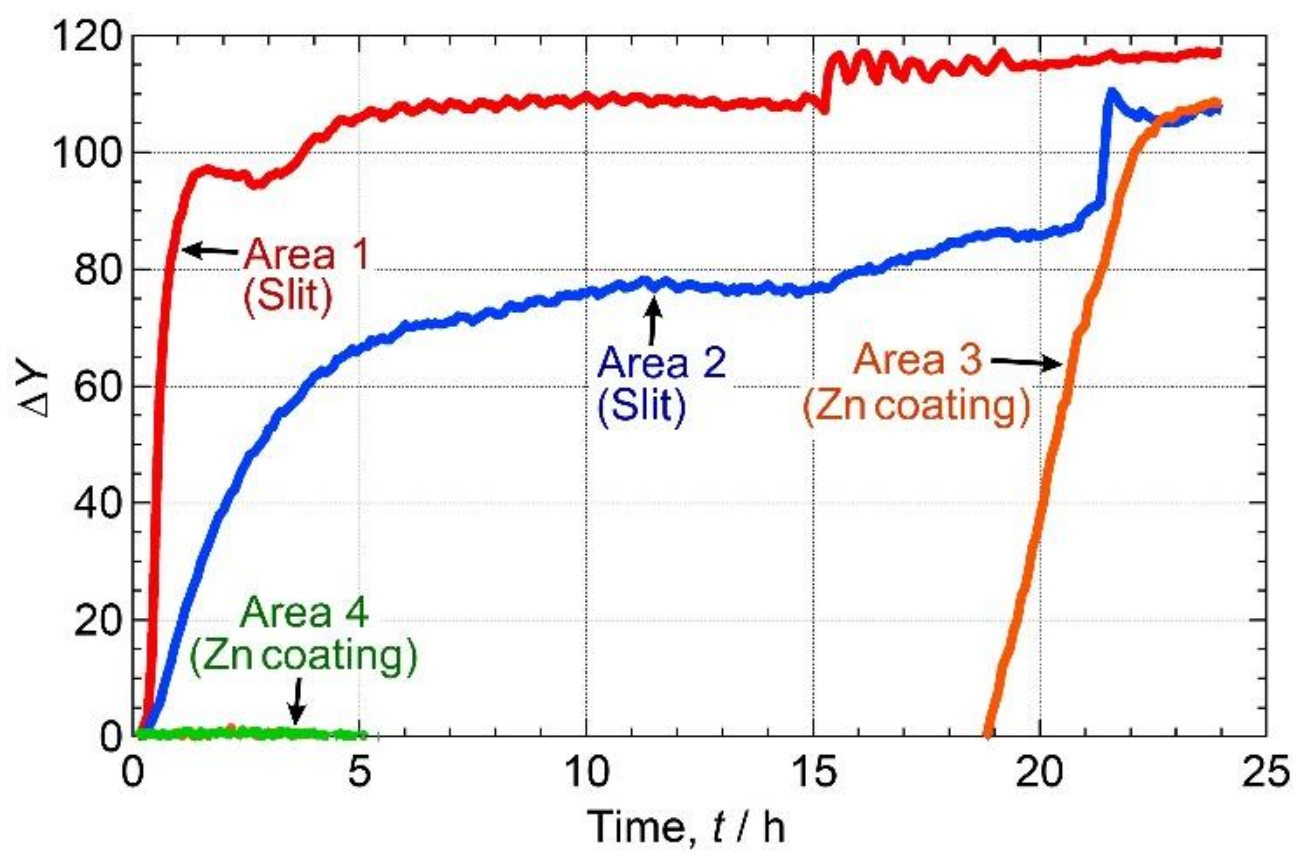

**Figure 8.** The time variations of the average $\Delta Y$ value in Areas 1−4, as indicated by the red squares in Figure 7.

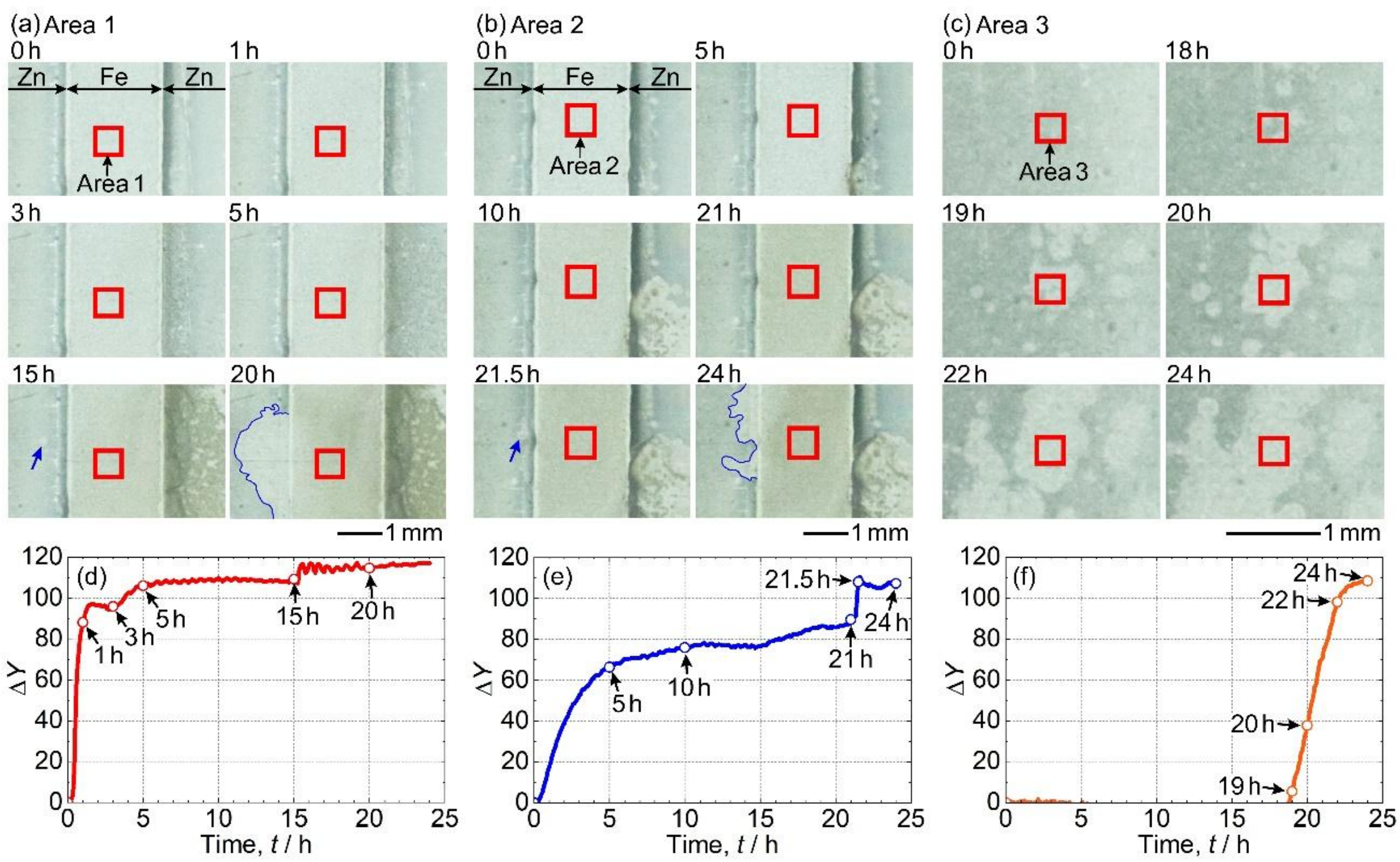

**Figure 9.** Optical images of (a) Area 1, (b) Area 2, and (c) Area 3 and their surrounding regions, as shown in Figure 7. The average $\Delta Y$ values in Areas 1−3 and the time points of optical images acquisition are indicated by the solid curves and white dots in (d)−(f), respectively.

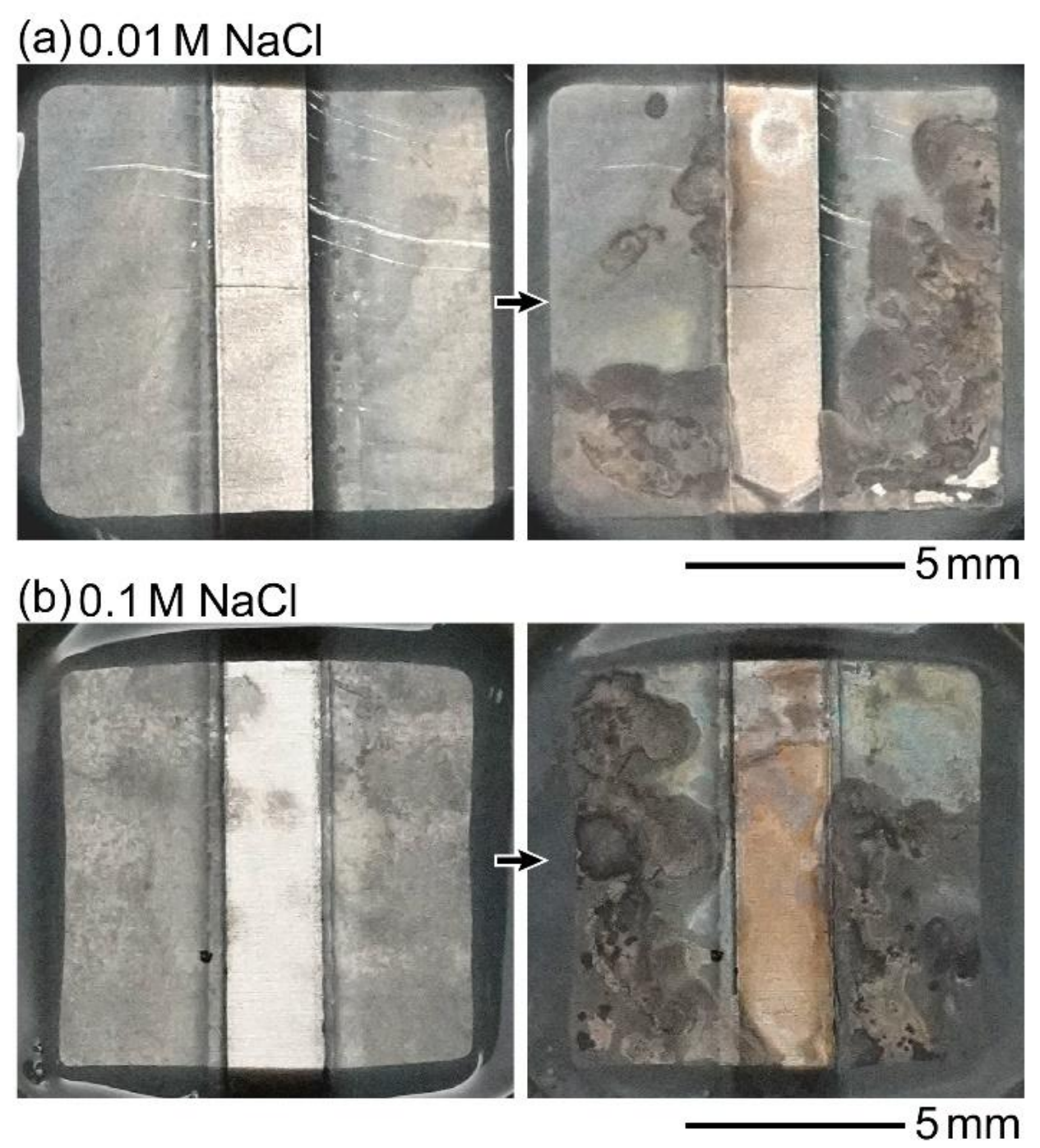

**Figure 10.** Optical images of the hydrogen entry side before (left) and after (right) the corrosion test in (a) 0.01 and (b) 0.1 M NaCl, corresponding to the Zn-coated steel shown in Figure 4b.

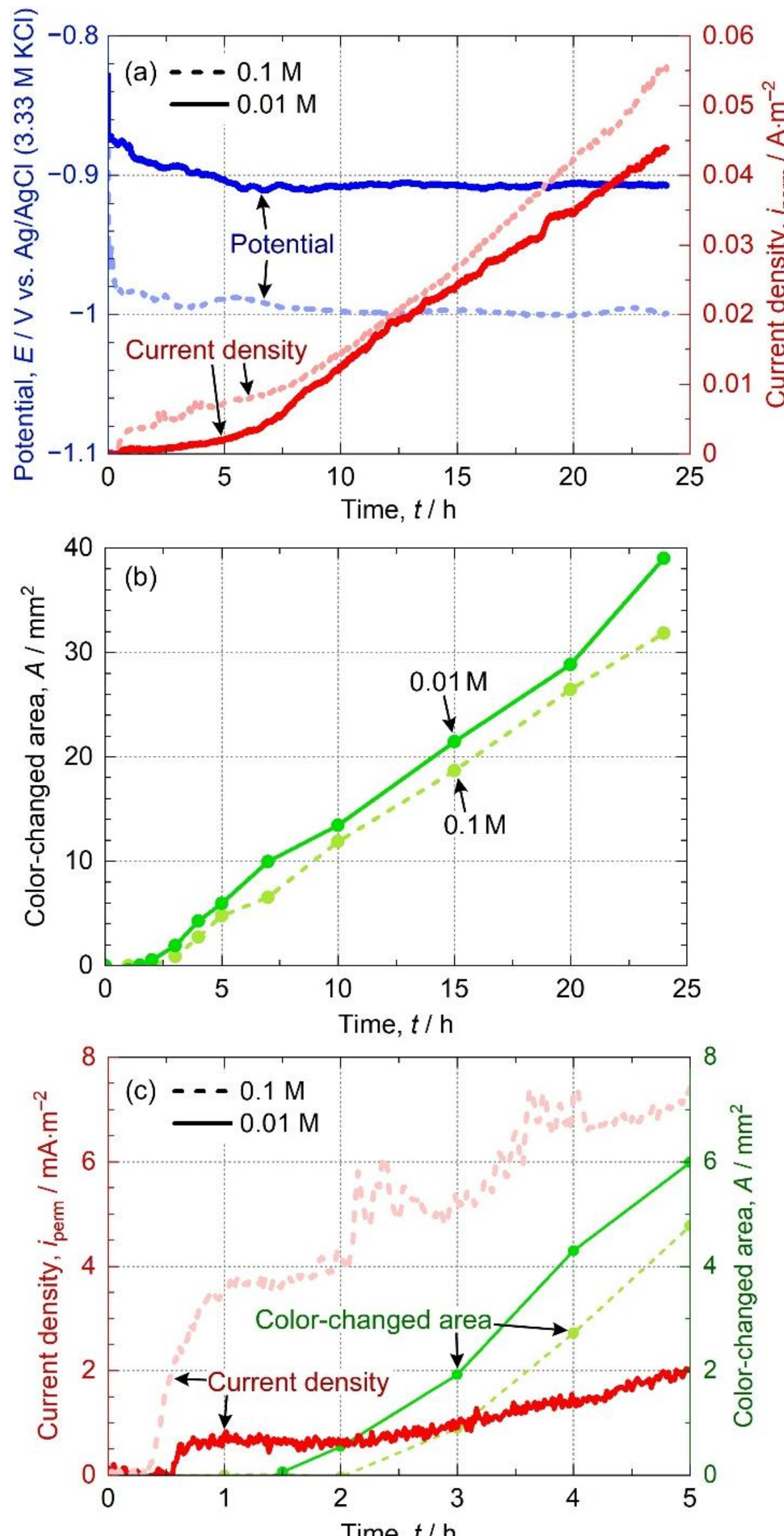

**Figure 11.** (a) Time variations of OCP (blue) and $i_{perm}$ (red) of the Zn-coated steel in 0.01 M NaCl. The results for 0.1 M NaCl from Figure 4c are shown as dotted curves for reference. (b) Time variations of the color-changed area of the Zn coating (green) during the corrosion test shown in (a). (c) Enlarged view of time variations of $i_{perm}$ and color-changed area presented in (a) and (b).

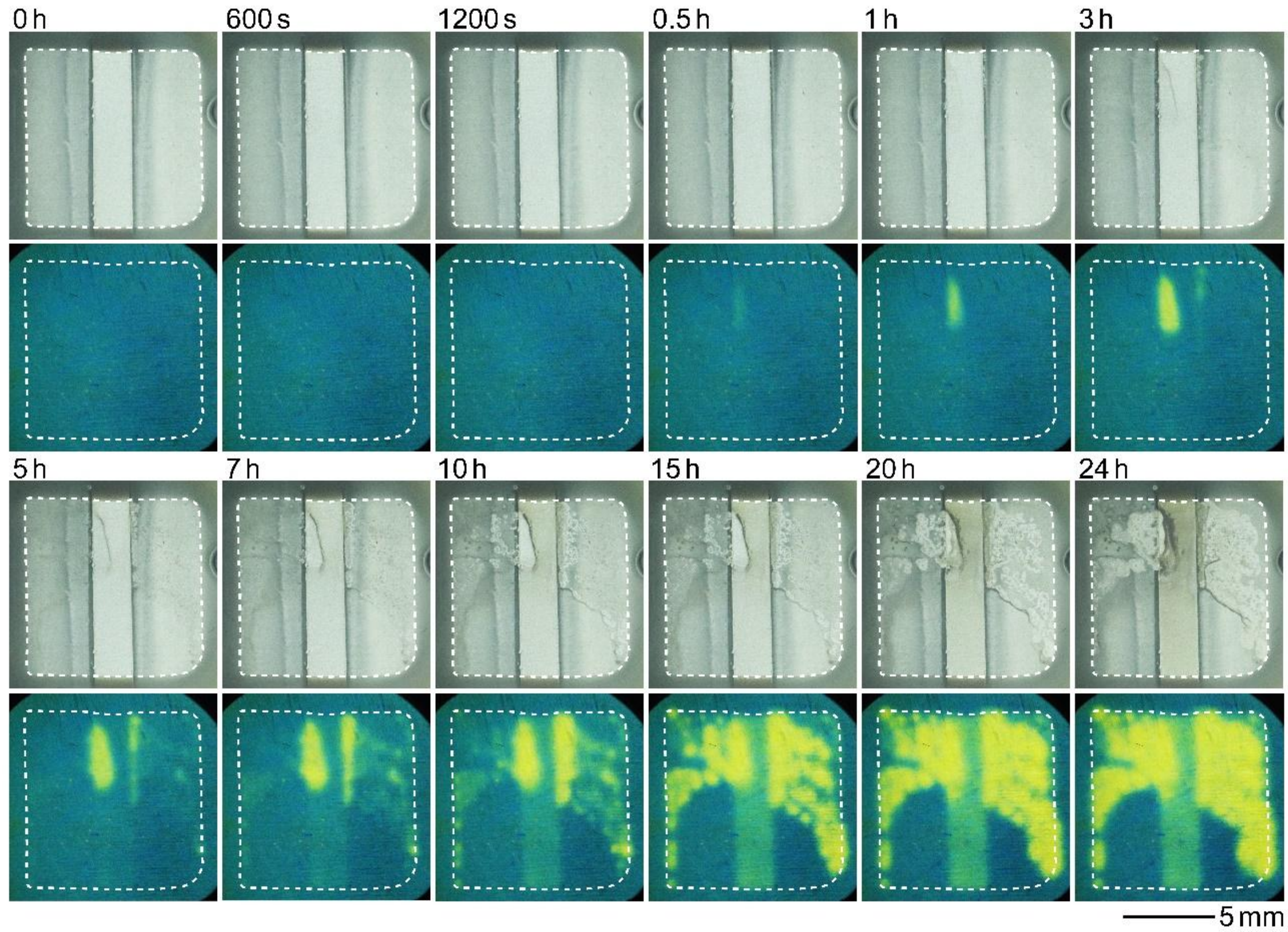

**Figure 12.** Optical images of the hydrogen entry (upper row) and detection (lower row) sides during the hydrogen visualization test in 0.01 M NaCl.

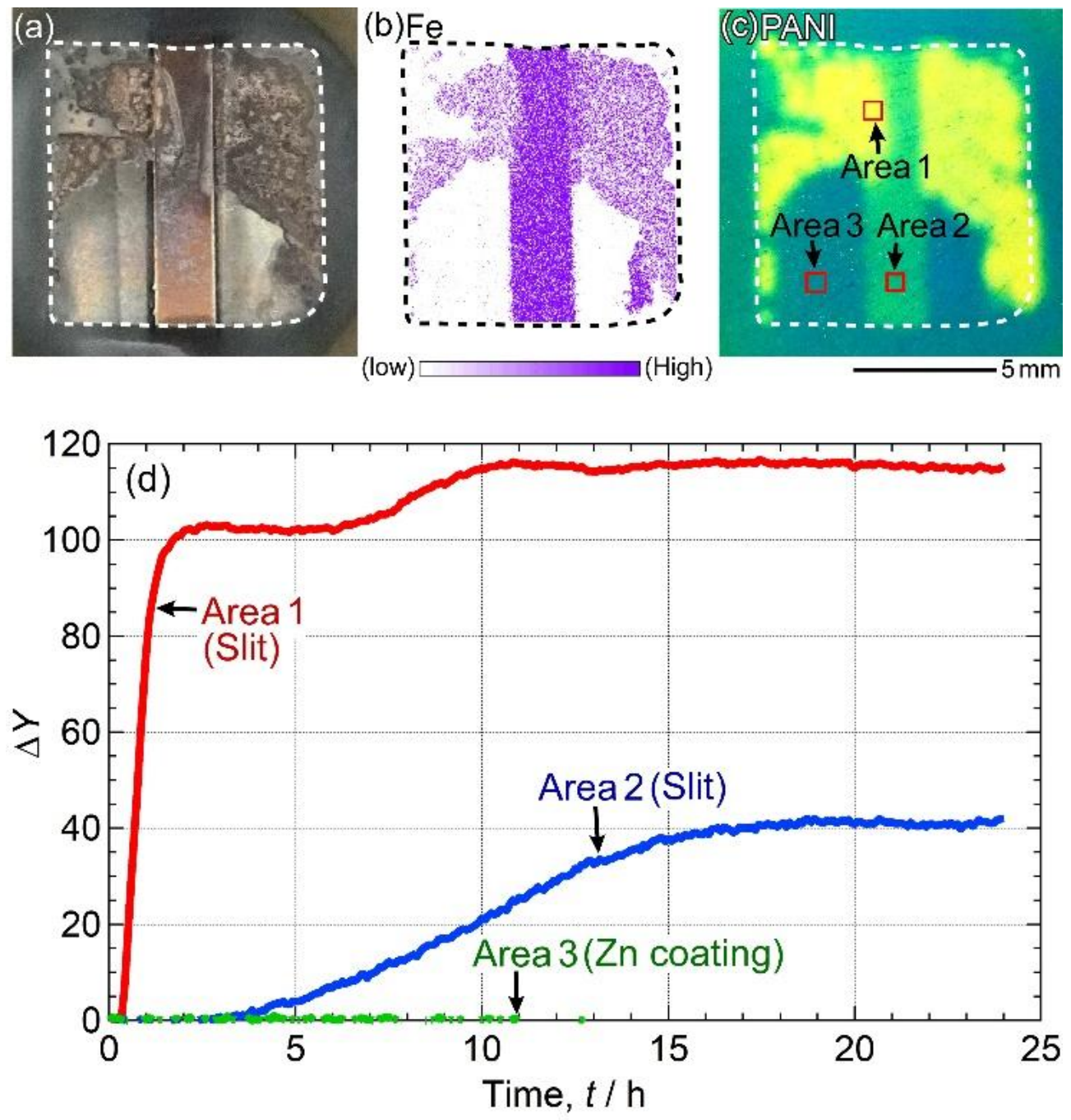

**Figure 13.** (a) Optical image and (b) elemental map of Fe on the hydrogen entry side, and (c) optical image of hydrogen detection side after the hydrogen visualization test in 0.01 M NaCl. (d) The time variations of the average $\Delta Y$ values in Areas 1–3, as indicated by the red squares in (c).

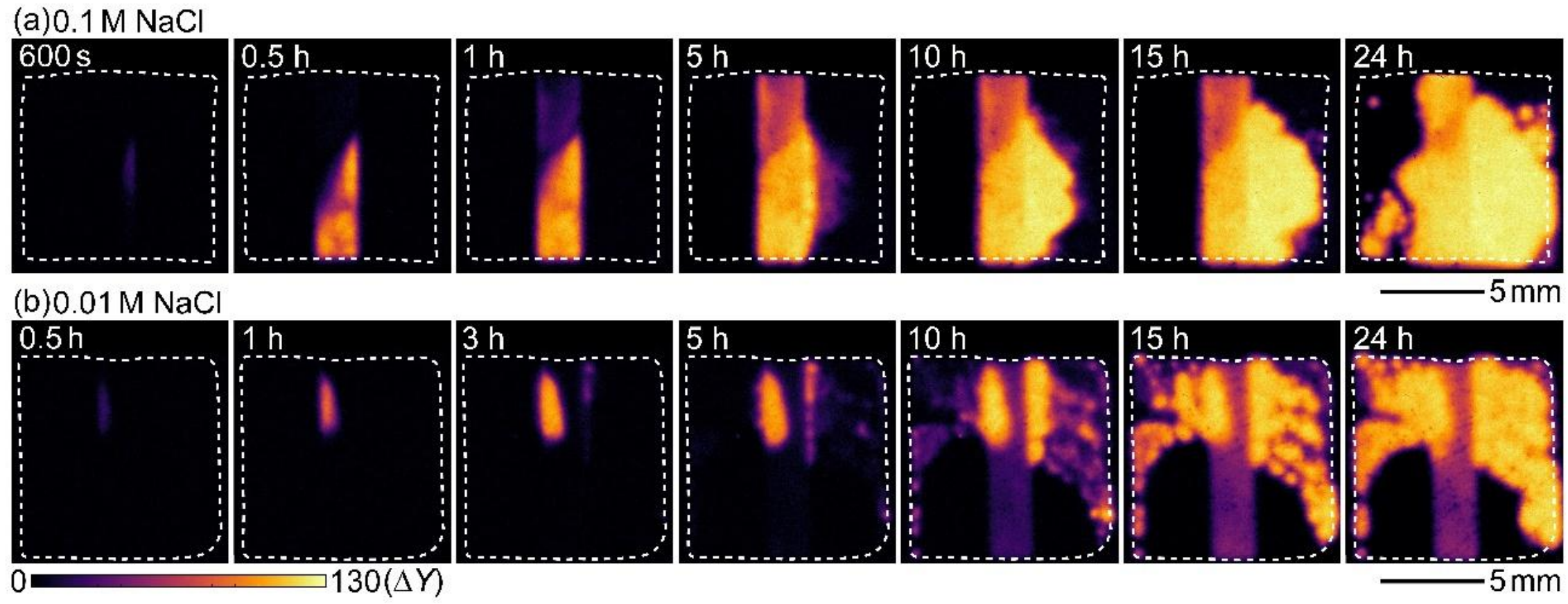

**Figure 14.** Contour maps of the $\Delta Y$ values of the hydrogen detection side shown in (a) Figure 7 and (b) Figure 12.

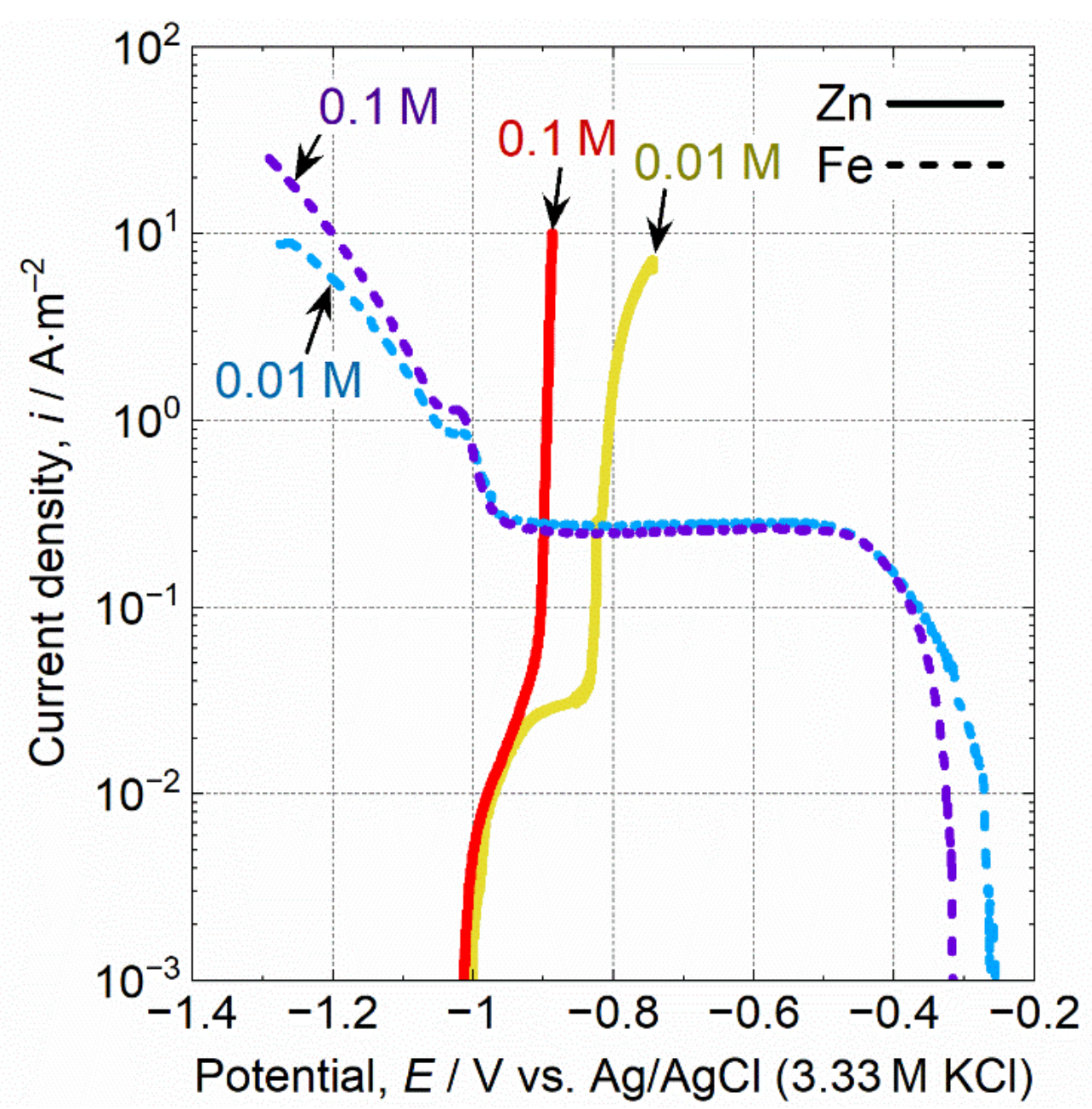

**Figure 15.** Anodic polarization curves of Zn plate and cathodic polarization curves of bare steel in 0.1 and 0.01 M NaCl.

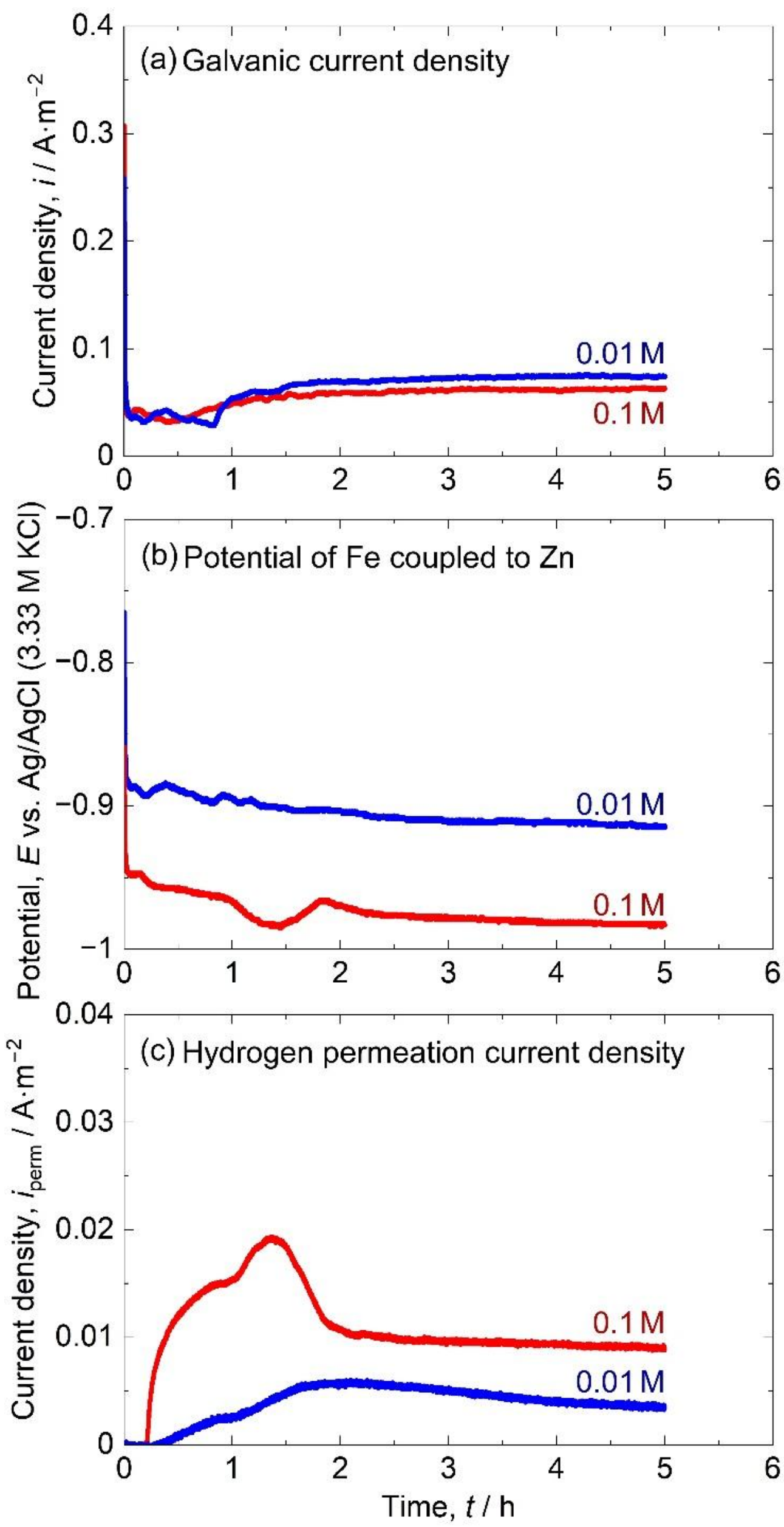

**Figure 16.** Time variations of (a) galvanic current density, (b) OCP, and (c) $i_{perm}$ of the bare steel coupled to Zn plate in 0.1 and 0.01 M NaCl.

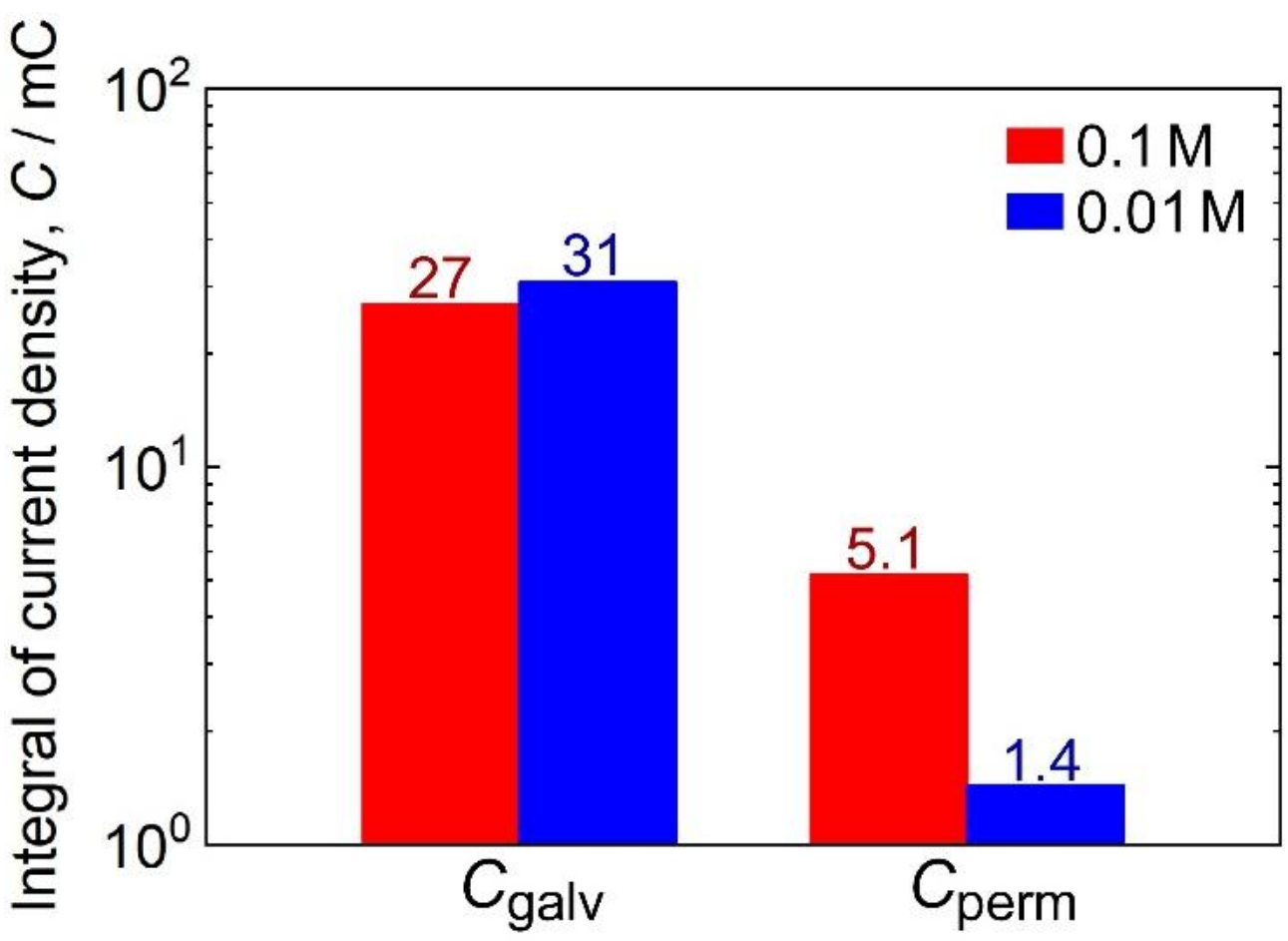

**Figure 17.** The integral values of galvanic current density ($C_{galv}$) and hydrogen permeation current density ($C_{perm}$), corresponding to the galvanic and hydrogen permeation current densities shown in Figures 16a and 16c.

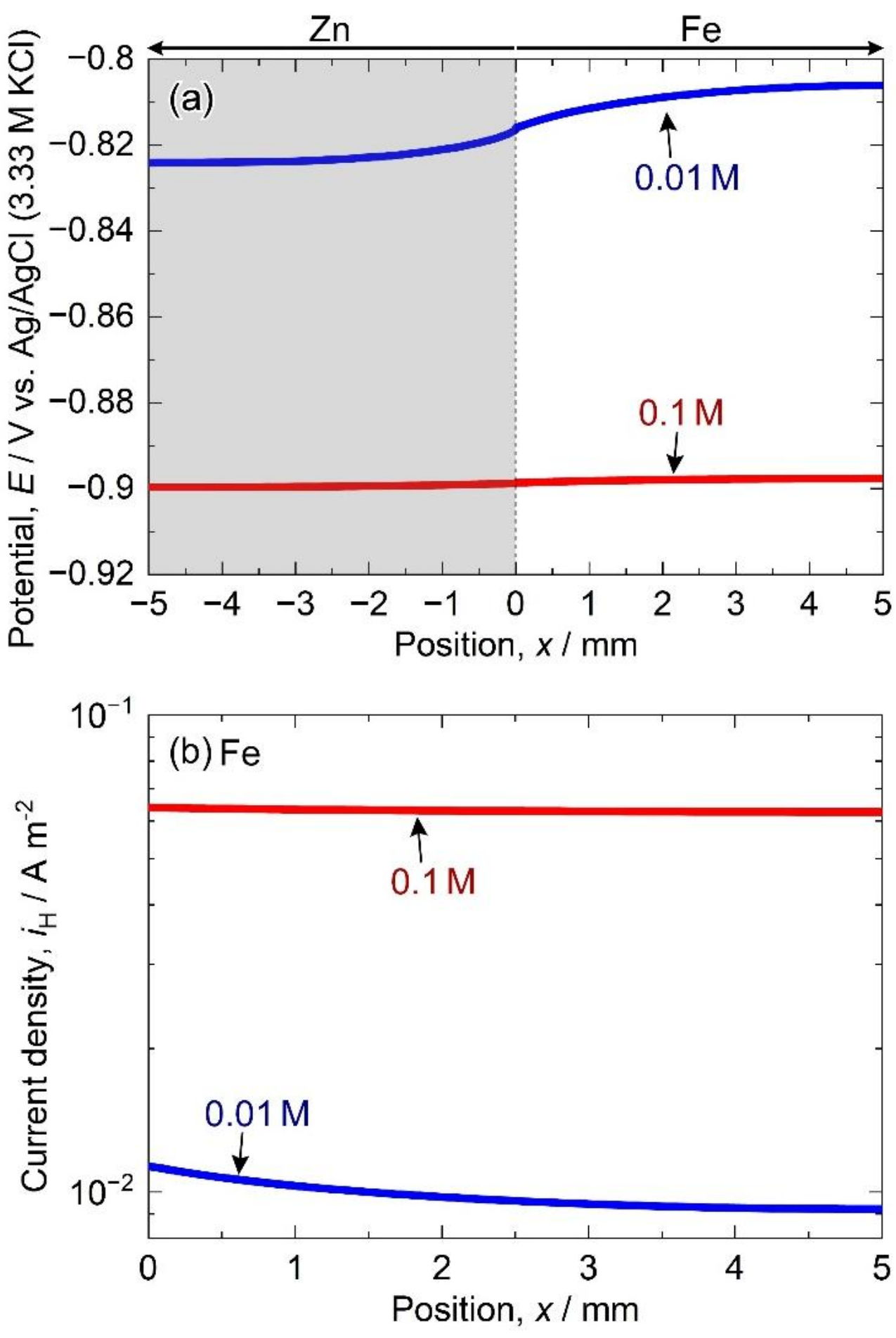

**Figure 18.** Distributions of (a) potential and (b) current density of HER on Zn-coated steels in 0.1 and 0.01 M NaCl.

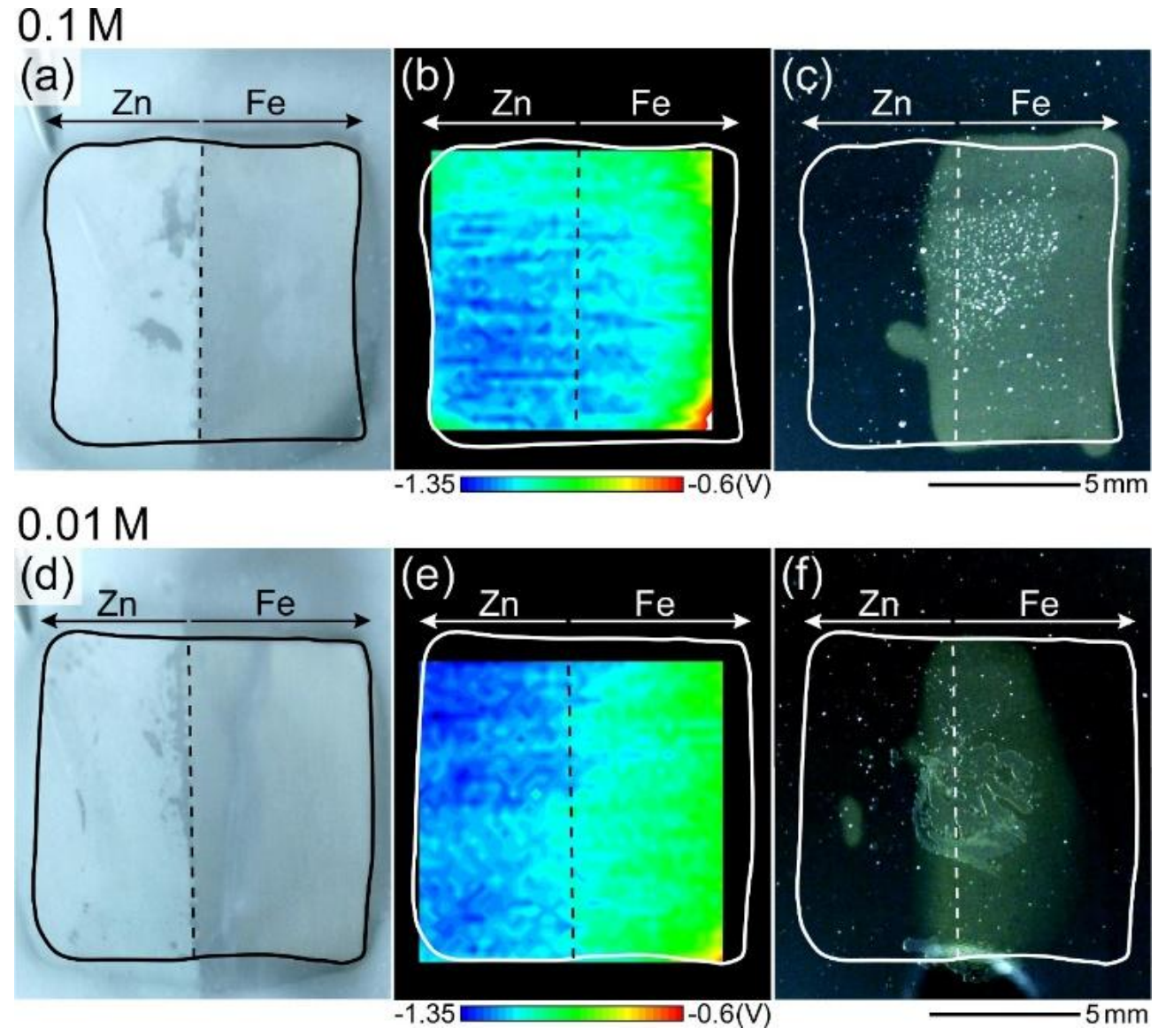

**Figure 19.** (a) Optical image and (b) potential distribution of the hydrogen entry side of the Zn-coated steel immersed in 0.1 M NaCl, and (c) corresponding hydrogen detection side after 7.2 h of immersion. (d) Optical image and (e) potential distribution of the hydrogen entry side of the Zn-coated steel immersed in 0.01 M NaCl, and (f) corresponding hydrogen detection side after 7.2 h of immersion.